\documentclass[useAMS,usenatbib,usegraphicx]{mn2e}
\usepackage{times}

\newif\ifAMStwofonts
\AMStwofontstrue

\newcommand{\simlt}{\lower.5ex\hbox{$\; \buildrel < \over \sim \;$}}
\newcommand{\simgt}{\lower.5ex\hbox{$\; \buildrel > \over \sim \;$}}
\newcommand{\be}{\begin{equation}}
\newcommand{\ba}{\begin{eqnarray}}
\newcommand{\ee}{\end{equation}}
\newcommand{\ea}{\end{eqnarray}}

\title[Early-type galaxy formation in GOODS]
{Exploring the formation of spheroidal galaxies out to z$\sim$1.5 in GOODS}
\author[I. Ferreras et~al.]
{Ignacio Ferreras$^{1}$\thanks{E-mail: ferreras@star.ucl.ac.uk},
Thorsten Lisker$^2$, Anna Pasquali$^3$ and Sugata Kaviraj$^{1,4}$\\
$^1$ Mullard Space Science Laboratory, Unversity College London, 
Holmbury St Mary, Dorking, Surrey RH5 6NT\\
$^2$ Astronomisches Rechen-Institut, Zentrum f\"ur Astronomie, Universit\"at 
Heidelberg, M\"onchhofstr. 12-14, D-69120 Heidelberg, Germany\\
$^3$ Max-Planck-Institut f\"ur Astronomie, Koenigstuhl 17, D-69117
Heidelberg, Germany\\ 
$^4$ Astrophysics subdepartment, The Denys Wilkinson Building, 
Keble Road, Oxford OX1 3RH}

\begin{document}
\date{Submited for publication in MNRAS}
\pagerange{\pageref{firstpage}--\pageref{lastpage}} \pubyear{2009}
\maketitle
\label{firstpage}

\begin{abstract}
The formation of massive spheroidal galaxies is studied on a visually
classified sample extracted from the ACS/{\sl HST} images of the GOODS
North and South fields, covering a total area of 360 arcmin$^2$.  The
sample size (910 galaxies brighter than $i_{\rm F775W}=24$) allows us
to explore in detail the evolution over a wide range of redshifts
($0.4<z<1.5$; median redshift $0.68$).  Three key observables are
considered: comoving number density, internal colour distribution; and
the Kormendy relation.  The comoving number density of the most
massive galaxies is found not to change significantly with
redshift. Extrapolation of our sample to z=0 gives an increase of the
comoving number density of $M_s>10^{11}M_\odot$ galaxies by a factor 2
between z=1 and z=0, in contrast with a factor $\sim 50$ for lower
mass galaxies ($10^{10}<M_s/M_\odot<10^{11}$).  One quarter of the
whole sample of early-types are photometrically classified as blue
galaxies. On a volume-limited sample out to z$<$0.7, the average
stellar mass of the blue ellipticals is $5\cdot 10^9 M_\odot$ compared
to $4\cdot 10^{10}M_\odot$ for red ellipticals. On a volume-limited
subsample out to z=1.4 probing the brightest galaxies ($M_V<-21$) we
find the median redshift of blue and red early-types: 1.10 and 0.85,
respectively. Blue early-types only amount to 4\% of this sample
(compared to 26\% in the full sample).  The intrinsic colour
distribution correlates overall bluer colours with {\sl blue cores}
(positive radial gradients of colour), suggesting an inside-out
process of formation. The redshift evolution of the observed colour
gradients is incompatible with a significant variaton in stellar age
within each galaxy.  The slope of the Kormendy relation in the
subsample of massive galaxies does not change over $0.4<z<1.4$ and is
compatible with z=0 values. The ``zero point'' of the Kormendy
Relation (i.e. the surface brightness at a fixed half-light radius) is
1~mag fainter (in the $B$ band) for the subsample of low-mass 
(M$_s<3.5\cdot 10^{10}$M$_\odot$) early-types.
\end{abstract}


\section{Introduction}
For several decades the formation of luminous early-type galaxies
has occupied a central debate in observational astrophysics. A
comprehensive literature on these objects has convincingly
established that the bulk of the star formation in these systems
takes place at high redshift \citep[z$\simgt$ 2 e.g. ][]{ble92,sed98} 
with possibly no star formation activity
thereafter -- a model that is often referred to as the `monolithic
collapse hypothesis'. This is borne out primarily by the
\emph{optical} properties of early-type populations and their
strict obedience to simple scaling relations over a large range in
redshift. This includes the small scatter in the early-type
`Fundamental Plane' \citep[e.g. ][]{jor96, sag97}
and its lack of evolution with look-back time \citep[e.g. ][]{For98,vdk96}, 
the homogeneity and lack of
redshift evolution in their optical colours \citep{ble92,Ellis97,sed98,vdk00}
and evidence for short ($<1$ Gyr) star formation timescales in
these systems, deduced from the over-abundance of $\alpha$
elements\citep[e.g. ][]{Thomas99}.

While the monolithic hypothesis reproduces the optical properties of
early-type galaxies remarkably well, it does not sit comfortably
within the currently accepted $\Lambda$CDM galaxy formation paradigm,
in which early-type galaxies are thought to form through a
hierarchical buildup process. Semi-analytical models, that combine
simulations of hierarchical structure formation with simple recipes to
describe the baryonic physics (star formation, chemical enrichment and
feedback from the supernovae and active galactic nuclei) have
enjoyed broad success in reproducing several properties of the galaxy
population in the low redshift Universe \citep[see e.g. ][]{Cole00,Hatton03}. 

While the predicted bulk \emph{photometric} properties of early-type
galaxies have met with reasonable success
\citep[e.g. ][]{Kav05,deLucia06,Kav07,Kav08b}, difficulties have remained in
reproducing the short star formation timescales implied by the high
alpha-enhancement ratios. While improved prescriptions may produce
better agreement to the observed ratios (Pipino et~al. 2008, in preparation) 
this remains one of the most important
challenges to the current incarnation of semi-analytical models in
terms of reproducing the properties of early-type galaxies.

Nevertheless, the discovery, using rest-frame UV data, of widespread
low-level recent star formation (RSF) in early-type galaxies, both at
low redshift \citep{Kav07} and at intermediate redshifts
\citep{FS00,Kav08}, points towards a more extended star formation
history than previously envisaged. While much of the work that
underpins monolithic collapse was conducted in the nearby Universe,
early-type populations at higher redshift are rapidly becoming
available through deep optical surveys. These surveys offer an
unprecedented window into the star formation histories of early-type
galaxies over the last 8 billion years, crucial for understanding
their formation and evolution.

Surveys that combine ground-based photometry with imaging from the
WFPC2 and ACS cameras on board the \emph{Hubble Space Telescope} ({\sl
HST}) additionally provide the angular resolution 
(FWHM $\sim 0.1^{\prime\prime}$) to perform spatially resolved analyses of stellar
populations down to physical projected sizes of $\sim1$ kpc at
z$\sim$1 \citep[e.g. ][]{fl05}.  In agreement with their low-redshift
counterparts, recent studies that have exploited such early-type data
at intermediate redshifts have found that the bulk of the star
formation in these objects does indeed form at high redshift.
However, tell-tale signatures of the recent star formation found in
the UV studies are clearly visible. A significant fraction of
early-types (up to 20\%) exhibit blue cores that are characteristic of
recent star formation \citep{men01,fl05,ap06}.  Furthemore, spectroscopy
has revealed [OII] emission lines in a similar fraction of field
early-type galaxies \citep{schade99,treu02}.


We present in this paper an extension of the sample in \citet{fl05} --
which was restricted to CDFS -- to the full coverage of the ACS/{\sl
HST} images of the GOODS North and South fields. However, notice that
the sample presented here {\sl does not} apply the selection based on
the Kormendy relation in contrast to \citet{fl05}. The only constraint
applied on the sample selection is visual classification.  The depth
and superb angular resolution of the images enable us to accurately
determine sizes and total magnitudes (which translate into stellar
masses). We can determine in a robust way the {\sl internal} colour
distribution and the Kormendy relation, out to redshifts z$\simlt$
1.2. We improve on \citet{fl05} by constraining the selection method
to visual classification (i.e. removing the Kormendy Relation
constraint) and by selecting volume-limited subsamples.

\section{The sample}

We describe in this section the definition of our $i$-band selected
(F775W), visually classified sample of early-type galaxies.  The
sample is extracted from the ACS/{\sl HST} images of the Great
Observatories Origins Deep Survey \citep[GOODS; ][]{goods}. Both North
and South fields are included, covering a total area of 320~arcmin$^2$
equally split between the two fields. The images consist of four deep
exposures through ACS passbands F435W ($B$), F606W ($V$), F775W ($i$)
and F850LP ($z$). The images are drizzled with a final pixel size of
30~mas and include a weight map used for the estimates of photometric
uncertainties. The weight maps also allow us to generate realistic
mock galaxies used to assess the accuracy of the half-light radii (see
appendix~A). The visual classification was performed on the version
1.0 images, whereas all the subsequent work: photometry, size
determination and intrinsic colour distribution was done on the
recently released version 2.0 images (Giavalisco and the GOODS Team,
in preparation). This version has a significant increase in the S/N of
the $z$ and the $i$ bands. The methodology presented in this paper was
already applied to the v1.0 images. We find an improvement on the
photometric and size uncertainties, but no major difference in the
overall behaviour of the sample.

\subsection{Size and Magnitude Estimation}

We started by selecting all objects with magnitude ${\rm
MAG\_AUTO}<24.0$ from the GOODS ACS $i_{F775W}$ source catalogues (v1.1;
detections based on the $z_{F850LP}$ band).
\footnote{http://archive.stsci.edu/pub/hlsp/goods/} 
We use hereafter elliptical apertures for each galaxy, adopting the
source centre and ellipse shape (position angle and ellipticity) as
given in the catalogues, keeping it fixed throughout the
process. Several objects missed by the GOODS detection,
presumably due to proximity of a bright neighbouring source, were
inserted manually, using Source Extractor \citep{sex} to determine
their shape. This initial sample contains 7,462 objects (3,992 in
GOODS-North and 3,470 in GOODS-South).

The GOODS ACS images suffer from residual (i.e.\ non-zero) background
flux, which typically reaches a level of 1/5 of the noise RMS in the
$i_{F775W}$ band. This background, along with a value of the noise
RMS, was determined individually for each galaxy as a single value
that corresponds to the median pixel value within a $21''\times
21''$-box, applying five iterations of clipping outliers at 2.3
standard deviations. All sources were masked in this process, using
2.0 times the KRON\_RADIUS, taken from the official GOODS catalogues.

Proper masking of neighbouring galaxies is crucial for a reliable
determination of the Petrosian radius \citep{pet76}. On the one hand,
object masks should be large enough to cover all contaminating light
of close neighbours. On the other hand, if neighbour masks are too
large they could encompass a big part of the target galaxy, rendering
unreliable estimates of the Petrosian Radius.

After testing several ways to define the masks -- e.g.\ using the
KRON\_RADIUS or a preliminary Petrosian radius -- we adopted an
elliptical aperture that extends out to a surface brightness
corresponding to one half of the local noise RMS (``half-noise
semimajor axis'' or ``half-noise SMA''). In addition, we apply an
iterative outlier clipping procedure in the surface brightness
calculations used to determine the Petrosian radius, as described
below.  In the determination of the half-noise SMA, initial object
masks were used that were based on the KRON\_RADIUS from the catalogues.

For each galaxy, we define a ``Petrosian semimajor axis''
(hereafter Petrosian SMA, $a_{\rm Petro}$), i.e., in the calculation
of the Petrosian radius, we use ellipses instead of circles
\citep[cf.][]{lot04,p3}. As mentioned above, the elliptical shape and
source centre of each object were taken from the GOODS official
catalogues, and kept fixed during the process. The Petrosian SMA was
defined as the semimajor axis $a$ at which the local intensity falls
below one fifth of the average intensity within $a$. The local
intensity is measured as the average intensity within an elliptical
annulus reaching from $0.9 a$ to $1.1 a$, applying five iterations of
clipping outlying pixel values at 2.3 standard
deviations. Neighbouring objects were masked using the half-noise SMA,
or the KRON\_RADIUS for sources with ${\rm MAG\_AUTO}\ge 24.0$.  For
250 objects (3.4\%), a neighbouring source was so close that its mask
encompassed the centre of the target galaxy, rendering a Petrosian SMA
derivation impossible.  For another 72 objects (1.0\%), the derived
Petrosian SMA is clearly too large from visual inspection, in most
cases caused by diffuse light or background inhomogeneities.

To obtain the total magnitude, effective radius, and effective surface
brightness of each galaxy, we measured the $i_{F775W}$ flux within $a=
1.5\,a_{\rm Petro}$, and the semimajor axis containing 50\% and 90\%
of this flux ($a_{50}$ and $a_{90}$, respectively).  For galaxy images
affected by a neighbour mask -- therefore without a Petrosian SMA --
we used the KRON\_RADIUS instead of $a = 1.5\,a_{\rm Petro}$, and
performed no masking of neighbours.  For those whose Petrosian SMA was
found to be too large, we also adopted the KRON\_RADIUS, but did mask
neighbouring galaxies.  From a comparison of catalog magnitudes
(MAG\_AUTO) with final magnitudes ($m_i$) for our early types, we find
that for 3-4\% of the objects, the final magnitude is brighter by more
than $0.5$\,mag than the catalog magnitude. Based on our preselection
of galaxies with ${\rm MAG\_AUTO}< 24.0$, we estimate our completeness
to be 95\% at $m_i\le 23.5$\,mag.

\begin{figure}
\begin{center}
\includegraphics[width=3.5in]{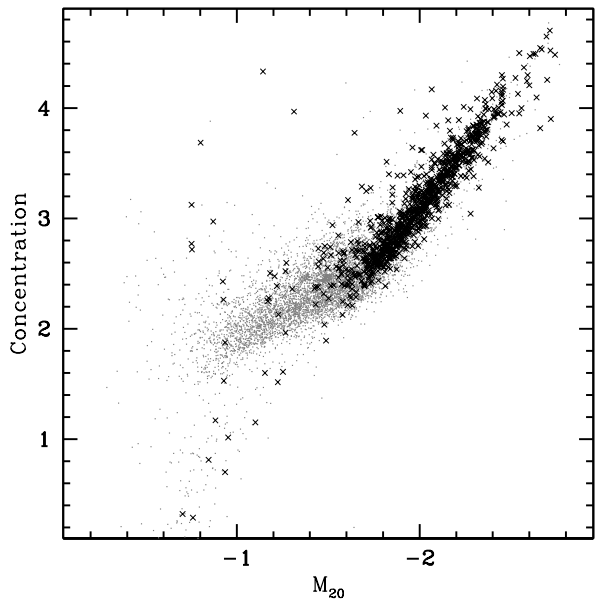}
\end{center}
\vskip-0.3truein
\caption{Distribution of concentration, as defined in \citet{ber00}
and M$_{20}$ for all galaxies with $i_{\rm F775W}\leq 24$ (grey
dots). Our final -- visually classified -- sample of early-type
galaxies in the GOODS North and South fields is shown as black crosses.
At the top-right corner C and M$_{20}$ correspond to a one-component distribution
with a high Sersic index. Towards the lower left corner, the contribution of a
second component (i.e. a disk) makes M$_{20}$ increase faster than C, resulting in 
a change of slope.
\label{fig:morph}}
\end{figure}

The flux lost due to masked areas within the aperture was corrected
by assigning the azimuthally averaged intensity to the masked
pixels.  We then applied a correction to the estimate of half-light
radius and apparent magnitude. This correction was obtained from
simulations of galaxies with Sersic surface brightness profiles and
realistic noise obtained from the GOODS/ACS weight maps. Notice that
this correction fixes two different effects: the dominant one in this
case comes from the fact that the apparent sizes are close to the
resolution limit of the observations. The second correction is
required as light measured within an aperture will lose flux, and this
loss is related to the steepness of the surface brightness profile
\citep[see e.g. ][]{gra05a, gra05b}. Our simulations show that the
correction can be written as a function of two observables: the
dominant one is the estimated R$_e$ and the second order one is the
concentration, measured as R$_{90}/$R$_{50}$ (see appendix~A for
details). 

The simulations allow us to assess the accuracy of the retrieved size,
magnitude and apparent surface brightness.  Four mock samples of 910
galaxies each were generated with the same values of $i_{\rm F775W}$
and R$_e$ as the real sample. These galaxies were given a random
eccentricity $0<e<0.6$, position angle, and \citet{ser68} index
$2<n<4$. We could sucessfully retrieve the
half-light radius with an accuracy of 9\%, the $i_{\rm F775W}$-band
{\sl total} apparent magnitude within 0.05~mag, and the average
$i_{\rm F775W}$-band surface brightness within the half-light radius
to within 0.16~mag/arcsec$^2$ (see appendix~A for details).

\begin{figure}
\begin{center}
\includegraphics[width=3.5in]{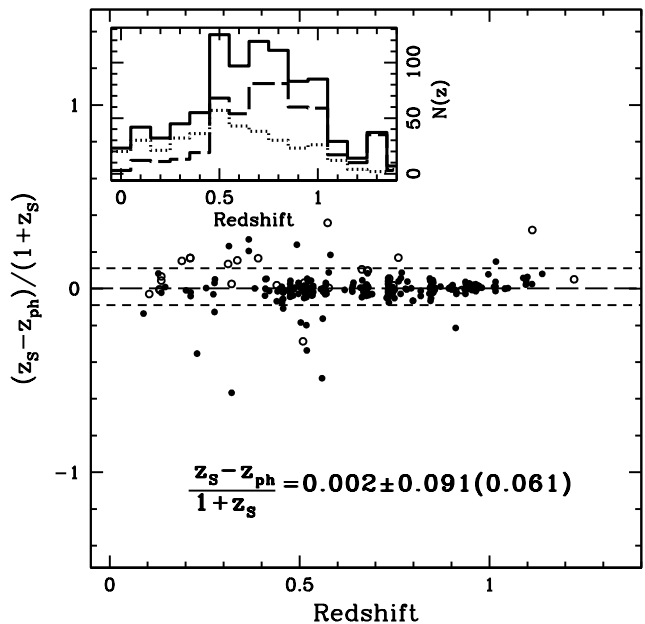}
\end{center}
\vskip-0.3truein
\caption{Comparison between our photometric redshifts and published
spectroscopic redshifts (see text for details). Solid (open) circles
correspond to photometric early (late) types. The comparison is
restricted to galaxies with $i_{\rm F775W}<23$ (with the full sample down
to $i_{\rm F775W}<24$, the standard deviation of the comparison is
0.13). The long (short) dashed lines show the sample mean (rms).  The
rms given in brackets correspond to the scatter when 3$\sigma$
outliers are removed from the sample.  Inset: Redshift distribution of
the complete sample (both North and South fields; thick solid line),
using spectroscopic redshifts where available. The other histograms
split the sample into photometric early- (dashed line) and late-types
(dotted line).
\label{fig:photz}}
\end{figure}

\subsection{Morphological Classification}

Galaxy morphology studies \citep[e.g. ][]{abr94,lot04,men04,sca07}
still lack a reliable classification method in an unsupervised way
with acceptable levels of contamination and completeness. Currently,
all these methods always require visual inspection. We decided to
tackle the selection of early-types only by visual inspection.  The
South field sample of early-types was taken from \citet{fl05}, with a
total of 377 early-type galaxies, out of a total of 3,470 galaxies at
$i_{\rm F775W}<24$ (10.9\%).  For the North field we follow an
identical procedure. All sources brighter than $i_{\rm F775W}=24$\,mag
are inspected by the four authors of this paper. We use {\sl
ds9}\footnote{http://hea-www.harvard.edu/RD/ds9/} to view each galaxy
in all four bands. After a first selection, we accept an early-type
galaxy if classified as such by at least three of us. A second pass of visual
inspection by the same people follows, this time restricted to the
list of galaxies already labelled during the first pass as early-types
by one or two of us. Galaxies classified as early-types by three of us
in this second pass are added to the final sample. Out of 3,992
sources in the North field, we selected 533 galaxies (13.4\%), with a
total in both fields of 910 galaxies. The different fraction is
compatible within Poisson statistics. Furthemore, Cosmic Variance
will certainly have an effect on the discrepancy between the
North and South fields.

Two ``non-parametric'' measures of galaxy morphology are shown in
Fig.~\ref{fig:morph} for our sample: the concentration index
\citep[$C$][]{ber00}, and the $M_{20}$ parameter
\citep{lot04}. $C=5\log R_{80}/R_{20}$, where $R_X$ is defined as the
radius within which one has $X$ percent of the total flux.  $M_{20}$
is the normalized second-order moment of the brightest 20\% of galaxy
pixels ({\sl no matter where they are}), and thus quantifies whether
these pixels are confined to a small region or widely spread. The figure,
analogous to Fig.~1 in \citet{fl05}, shows that the majority of
galaxies follow a well-defined relation in the $C$-$M_{20}$-plane,
which consists of two linear parts with different slope.  The change
in slope marks the transition between bulge-dominated and
disk-dominated or irregular objects: for the former, the brightest
20\% of pixels are fully located in the centre, whereas for the
latter, a growing fraction is located outside of the central
region.

\begin{figure}
\begin{center}
\includegraphics[width=3.5in]{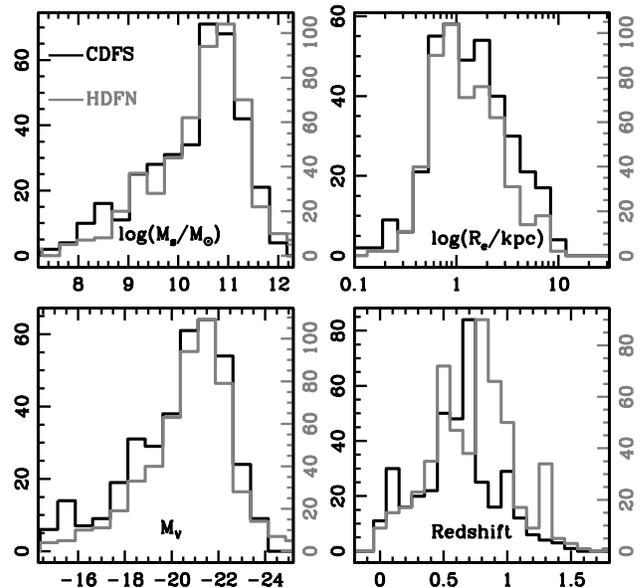}
\end{center}
\vskip-0.3truein
\caption{Distribution of various observables. The histograms
separate between the North (grey) and the South (black) fields in order
to illustrate the effect of Cosmic Variance.
\label{fig:distrib}}
\end{figure}

\subsection{Redshift and Rest-frame Estimates}

We retrieved the latest publicly available spectroscopic redshift
catalogues from the GOODS fields: FORS2 \citep{fors2z} and VVDS
\citep{vvdsz} in CDFS and the Team Keck Redshift Survey\citep{tkrs04}
in HDFN. In \citet{fl05} photometric redshifts were taken from
\citet{mob04} for galaxies without a spectroscopic measurement. In
this paper we determine our own photometric redshifts both in the
North and South fields for consistency. Our code is based on a 
template fitting technique without any priors and uses a set of eight
templates built from \citet{bc03} population synthesis models. Each
template is a $\tau$-model, i.e. a composite stellar population at
solar metallicity with an exponentially decaying star formation rate
($\tau$). We fix the formation redshift z$_{\rm F}$=3 for all
templates, with timescales: $\tau = \{ 0.05, 0.5, 1.0, 2.0, 4.0, 5.0,
6.0, 8.0\}$ Gyr.  We label these star formation histories with an
integer t=$\{0,\cdots,7\}$.  The templates are tranformed according to
a \citet{Fitz99} law for Galactic reddening with a colour excess
E(B-V)=0.012 (HDFN) and 0.010 (CDFS), values taken from \citet{sch98}
for the central positions of both fields. Each model is run over a
range of redshifts out to z=3 with the available photometry
estimated over a fixed aperture. The photometry comes from the
GOODS/ACS images ($B$,$V$,$i$,$z$) and is directly extracted from the
publicly available v2.0 images. For HDFN we also include ground-based
photometry from \citet{cap04}; which includes KPNO/4m $U$-band
and UH/2.2m $HK^\prime$ data. In CDFS optical and NIR
photometry is taken from WFI/NTT and ISAAC/VLT as used in the
photometric redshift catalogue of \citet{mob04}.

\begin{figure*}
\begin{center}
\includegraphics[width=3.2in]{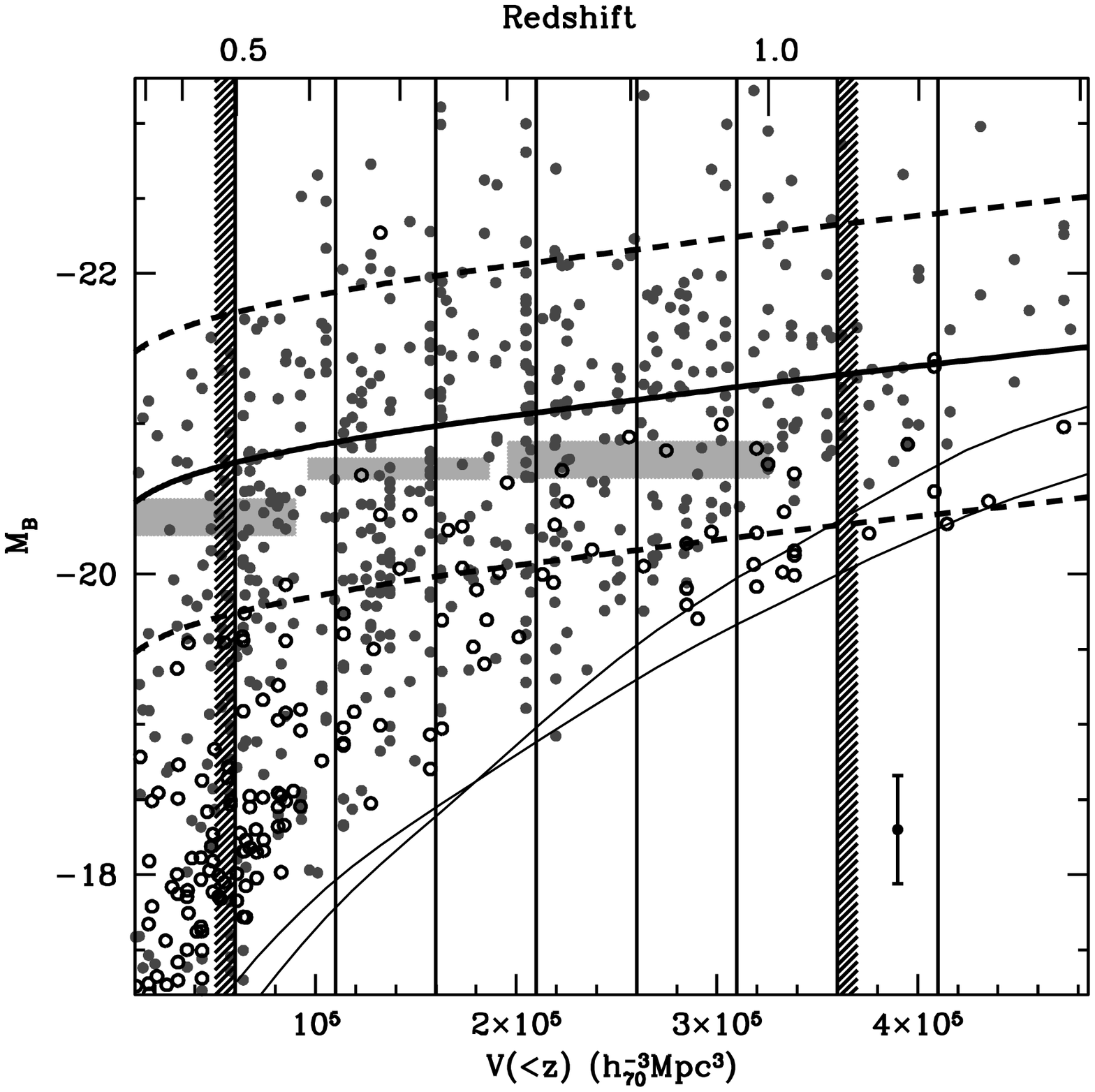}
\includegraphics[width=3.2in]{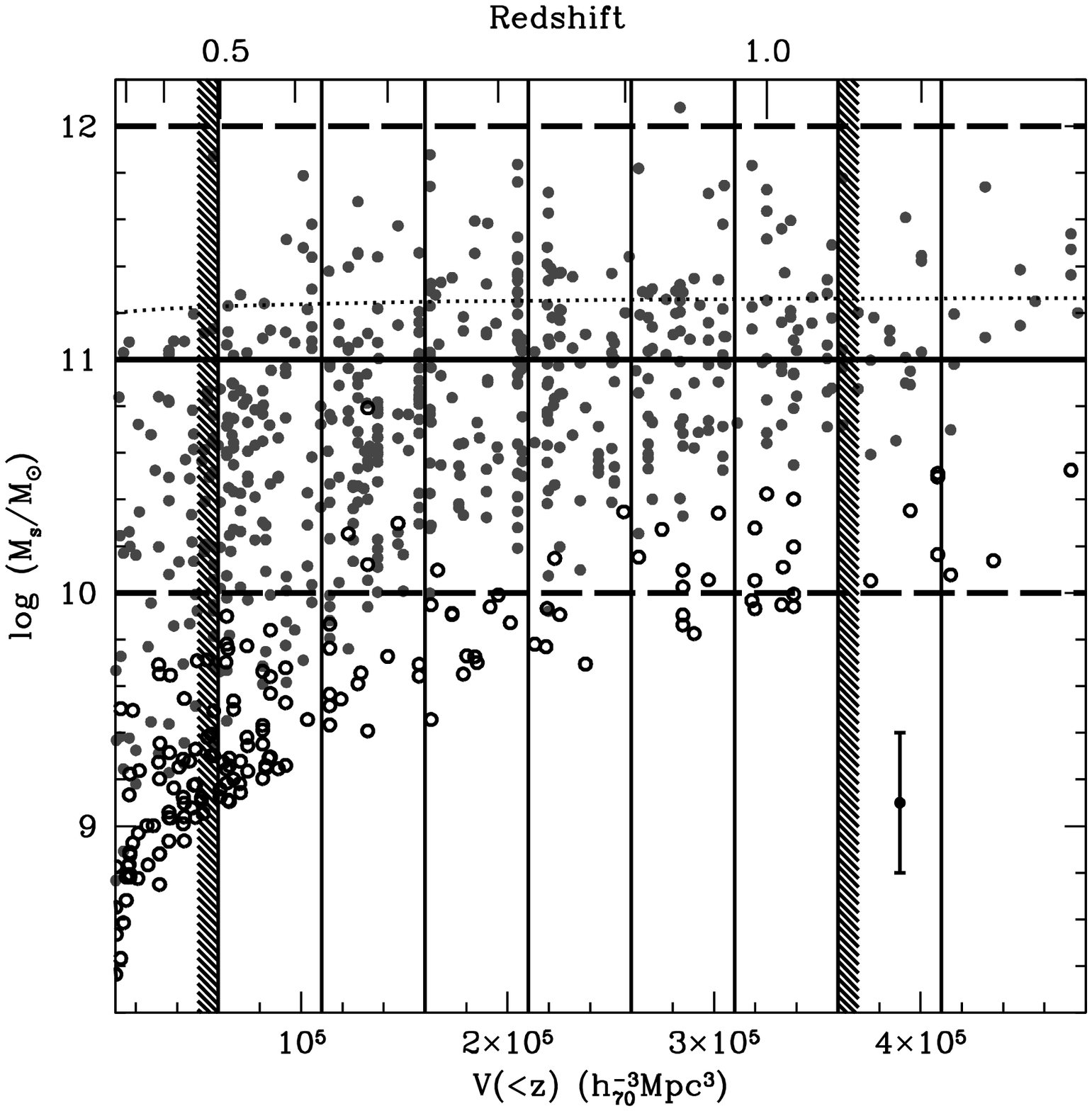}
\end{center}
\caption{Redshift evolution of the comoving number density: The
$B$-band absolute magnitude ({\sl left}) and stellar mass ({\sl
right}) is compared with $V(<z)$, i.e. the comoving volume within a
given redshift (the redshift appears in the top axis). Vertical lines
delimit redshift bins used in the analysis of the number density in
the next figure, with the upper and lower limits shown as shaded
regions. Galaxies are shown as grey dots (photometric early-type) or
open circles (late-type). The curved, thin solid lines in the left
panel are the limiting $i_{\rm F775W}=24$ for a photometric early- or
late-type. The thick line and dashed lines above and below illustrate
the sample separation into ``bright'' and ``faint'' subsets (or high and
low stellar mass in the right panel; see text for details). 
Grey areas in the left panel are $L_*$ values from \citet{il06}. 
The dashed line in the mass plot follows $M_*$ from \citet{fon06}.
\label{fig:volz}}
\end{figure*}

Figure~\ref{fig:photz} shows a comparison of photometric and
spectroscopic redshifts for galaxies brighter than $i_{\rm F775W}<23$.
The average (0.002) and scatter (0.091) of the discrepancy -- measured as
$\Delta z/(1+z_S)$ -- is comparable to the best redshift estimates 
available \citep{mob04,mob07}. If outliers at the $3\sigma$ level are
removed from the sample, the scatter reduces to 0.061. The fraction of
outliers is 2.9\% with no significant dependence on galaxy type. 

\begin{figure}
\begin{center}
\includegraphics[width=3.5in]{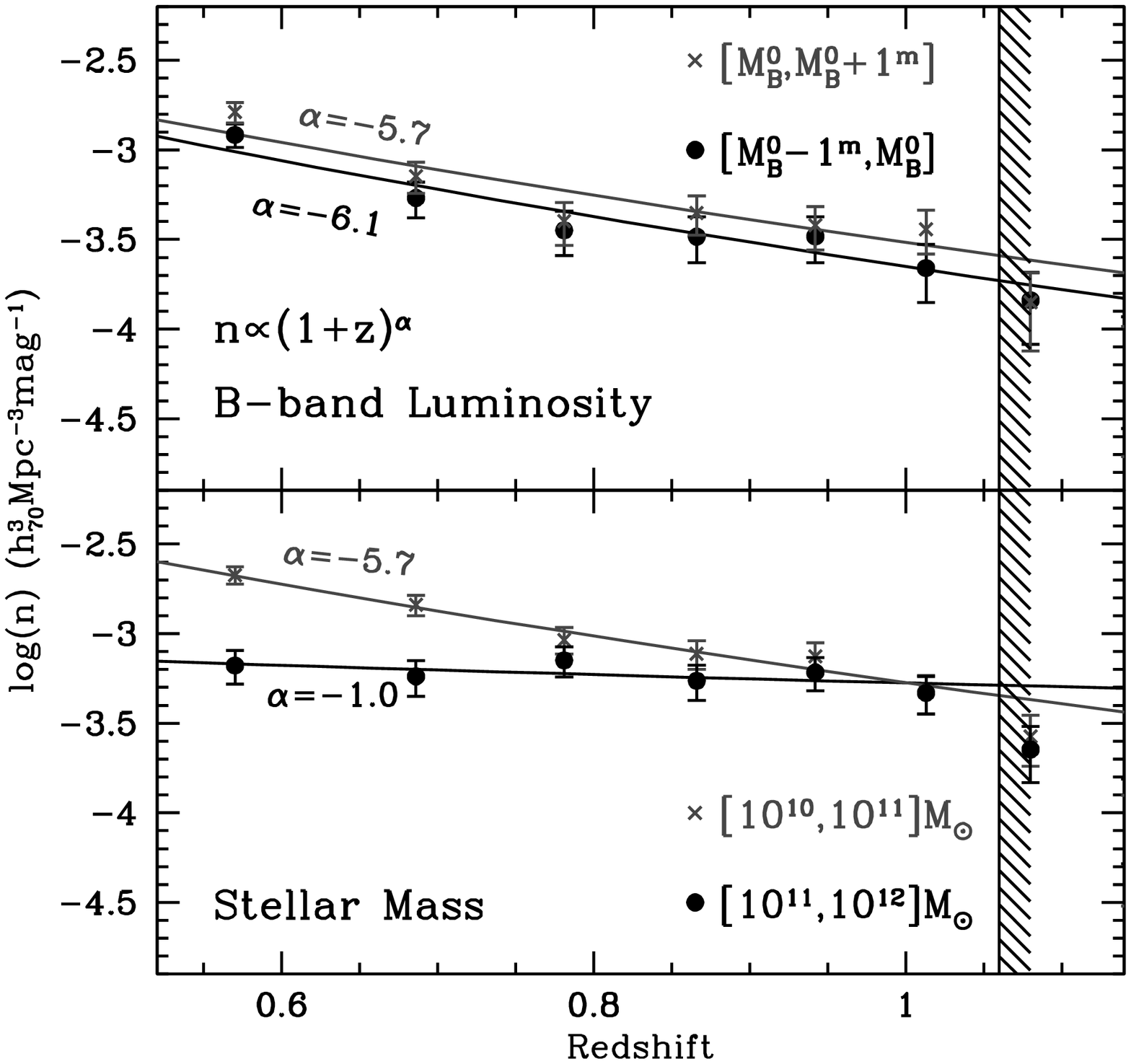}
\end{center}
\vskip-0.3truein
\caption{
Redshift evolution of the top end of the mass and luminosity
distribtution as illustrated in the previous figure. The black dots
correspond to the brighter (top panel) or more massive (bottom)
subsample. The fainter subsample is shown as grey crosses. Error bars
include Poisson noise and the uncertainty in the K-correction. The
lines give fits to a power law, with index $\alpha$, which is labelled
for all subsamples.
\label{fig:LFevol}}
\end{figure}

Our photometric redshift code also identifies each galaxy with the
template that gives the best fit. This allows us to split the sample
into ``photometric'' red and blue galaxies depending on whether the
best fit corresponds to old or younger stellar populations. We
compared this new classification -- using the templates presented
above -- with those done in \citet{fl05}, which were based on the
photometric types from \citet{mob04}. In this paper we choose the
separation between photometric red and blue ellipticals at the third
template (t=2): all galaxies which are best fit by an exponential timescale
of 1~Gyr or shorter are considered ``red'' throughout this paper. With
this choice, the number of ``red'' early-type galaxies in the total
sample is 670 ($\sim$75\%). The fraction of red early-types in the
North and South fields is 375/533 and 295/377, compatible within
Poisson error bars.

In addition, we used redshifts determined for a subsample of our
dataset, for which IF and AP have access to ACS/{\sl HST} slitless grism
spectroscopy (G800L) from the PEARS project (Probing Evolution And
Reionization Spectroscopically; PI S.~Malhotra; HST Proposal ID \# 10530)
The low-resolution spectra (R$\sim 50-80$) allowed us to secure
redshifts with comparable accuracy to the spectroscopic ones because
of the prominent 4000\AA\ break found in most of these galaxies. For
details we refer the interested reader to the paper on the analysis of
the grism data from the GOODS early-types (Ferreras et~al., in
preparation), or to a similar study on the HUDF \citep{ap06}.
Out of the total sample of 910 early-type
galaxies in this paper, we have spectroscopic or grism-based 
redshifts for 541 galaxies. Hence, 59.5\% of the total sample have
spectroscopically accurate redshifts.

The redshifts are used to determine rest-frame properties of the
galaxies. The physical projected half-light radius is determined by
using the angular diameter distance for a {\sl concordance} cosmology,
namely $\Omega_m=0.3$, $\Omega_\Lambda = 0.7$ and $H_0=72$\,km/s/Mpc.
The best fit template is compared to the apparent total magnitude to 
estimate the absolute magnitude and
stellar mass -- a \citet{chab03} Initial Mass Function is used
throughout this paper. From the ACS $B$ (where available), $V,i,z$
photometry we choose the passband with the lowest K-correction for the
transformation into absolute magnitude. As an estimate of the amount
of correction applied, the RMS of the K-corrections for the full
sample is 0.42~mag for M$_B$ and 0.27~mag for M$_V$. Regarding stellar
mass estimates, even though the age and metallicity distribution of a
composite stellar population cannot be constrained with broadband
photometry, the stellar mass can be reliably constrained to within
$0.2-0.3$~dex provided the adopted IMF gives an accurate
representation of the true initial mass function \cite[see
e.g.][]{fsb08}. In order to assess the effect of the passband on the
determination of stellar mass we compare our stellar masses -- based
on $i_{\rm F775W}$ -- with the $K_s$ photometry of the CDFS field from
ISAAC/{\sl VLT} (Retzlaff et~al., in preparation). Over the redshift range
probed by our sample, $i$-band and NIR mass estimates differ by less
than 0.15~dex (see appendix~B).

\begin{table}
\caption{The sample of early-type galaxies ($i_{\rm F775W}<24$)}
\label{tab:sample}
\begin{center}
\begin{tabular}{lccclcc}
\hline\hline
Sample & Number & $f_B^a$ & z$^b$ & M$_{\rm s}^b($M$_\odot)$ & $M_V^b$ & R$_e^b$(kpc)\\
\hline
HDFN     & 533 & 0.30 & 0.78 & $4.2\cdot 10^{10}$ & $-20.9$ & 1.10\\
CDFS     & 377 & 0.22 & 0.64 & $3.3\cdot 10^{10}$ & $-20.6$ & 1.26\\
All      & 910 & 0.26 & 0.68 & $3.7\cdot 10^{10}$ & $-20.8$ & 1.15\\
S1$^c$   & 411 & 0.04 & 0.85 & $1.2\cdot 10^{11}$ & $-21.9$ & 2.11\\
S2$^c$   & 280 & 0.11 & 0.56 & $3.4\cdot 10^{10}$ & $-20.6$ & 1.55\\
\hline\hline
\end{tabular}
\end{center}

$^a$ Blue early-type fraction: Red/Blue types selected from
photometric redshift best type (Red: t$\leq$ 2, i.e. star formation
timescale $\tau\leq 1$~Gyr, see text for details).
$^b$ Median value.
$^c$ S1 and S2 are volume-limited subsamples, defined in \S4.
\end{table}


Figure~\ref{fig:distrib} shows the distribution of stellar mass,
projected physical size, absolute magnitude and redshift. The sample
is shown separately for the North (grey) and South (black) fields.
Both fields have a similar range in mass or size, but the redshift
distribution illustrates the effect of cosmic variance over the
individual 160~arcmin$^2$ areas. The variation in the total number of
early-type galaxies (533 in the North vs. 377 in the South) is also
caused by such effect (the selection method being identical in both
fields).  Some of the basic properties of this sample is shown in
table~\ref{tab:sample}. The table also shows the properties of a
couple of volume-limited subsamples (S1 and S2) defined in \S4.

\section{Number density evolution: Luminosity vs. Mass}

A simple way to assess the redshift evolution in the number density of
early-type galaxies is illustrated in figure~\ref{fig:volz}, where the
$B$-band absolute magnitude ({\sl left}) or stellar mass ({\sl right})
is presented with respect to the comoving volume enclosed within the
redshift of each galaxy. Hence, counting galaxies within cells in this
figure gives the comoving number density at a given redshift. The thin
solid lines track the flux limit imposed by our search 
($i_{\rm AB}<24$). This limit depends on the K-correction adopted, thereby we
show in the figure the absolute magnitude limit taking the extrema of
the templates used in the photometric redshift analysis (i.e. types
t=0 and t=7).  Galaxies are shown according to the convention adopted
throughout this paper, namely red galaxies as grey solid dots
(photometric type t$\leq$ 2) and blue galaxies as black open circles
(t$>$2). In both cases -- absolute magnitude and stellar mass -- we
split the sample in two bins. In absolute magnitude we define 1~mag
bins (dashed lines) above and below a typical L$_\star$ galaxy
represented by the thick solid line, which corresponds to 
$M_B^0\sim -21$. In order to track its evolution with redshift, we assume an
exponential star formation history with formation redshift z$_{\rm F}$=5, 
timescale $\tau_{\rm SF}=0.5$~Gyr and solar metallicity.  For
comparison, the shaded boxes delimit the value of the $M_B^\star$
Schechter fits of the VIMOS-VLT Deep Survey for early-type galaxies
\citep{il06}. We decided to choose this binning criterion instead of a
fixed absolute magnitude in order to overcome the passive fading of
the stellar populations. The shaded areas show that our low-luminosity
bin roughly maps L$_*$ galaxies.

\begin{figure*}
\begin{center}
\includegraphics[width=3.2in]{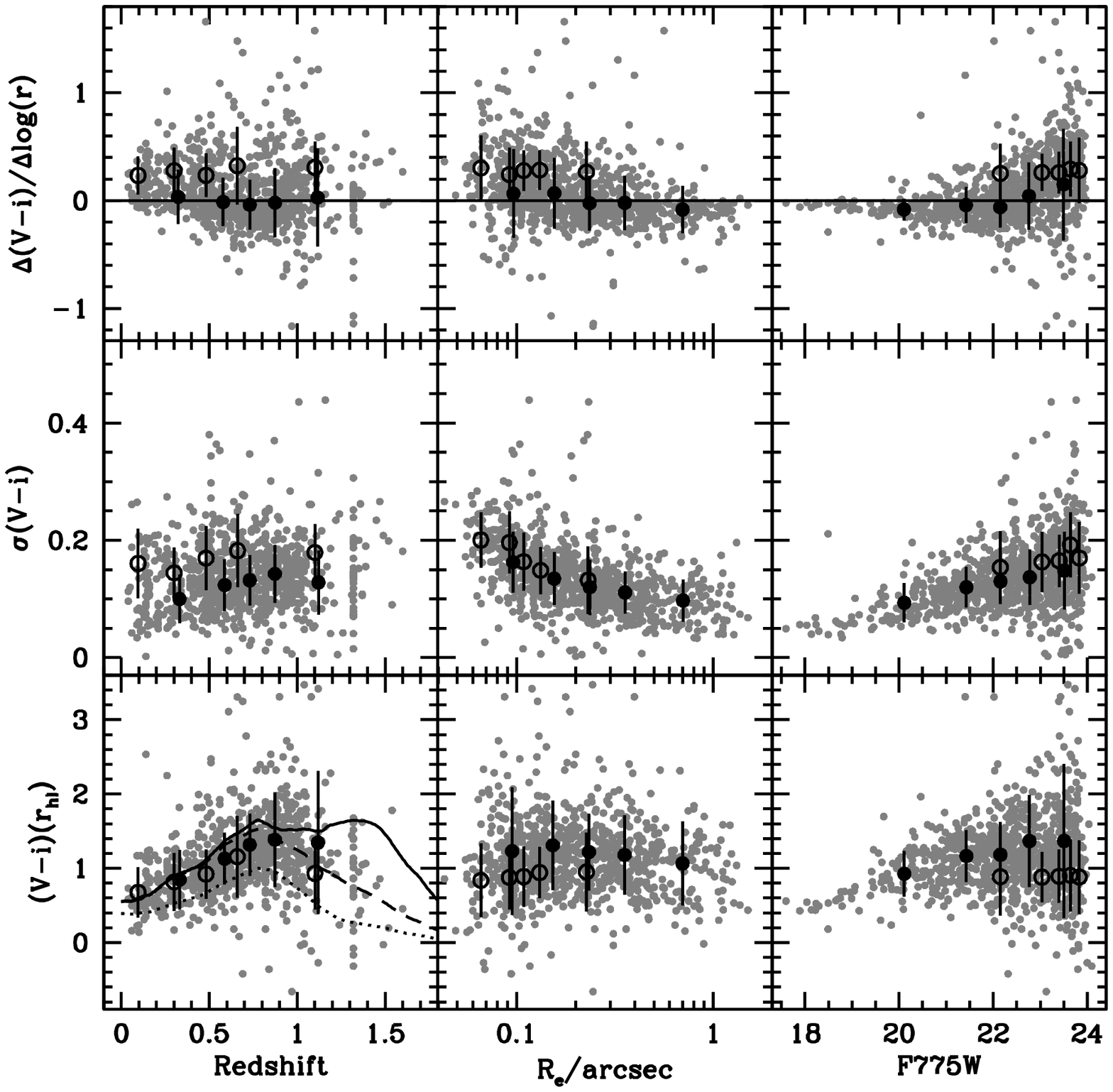}
\includegraphics[width=3.2in]{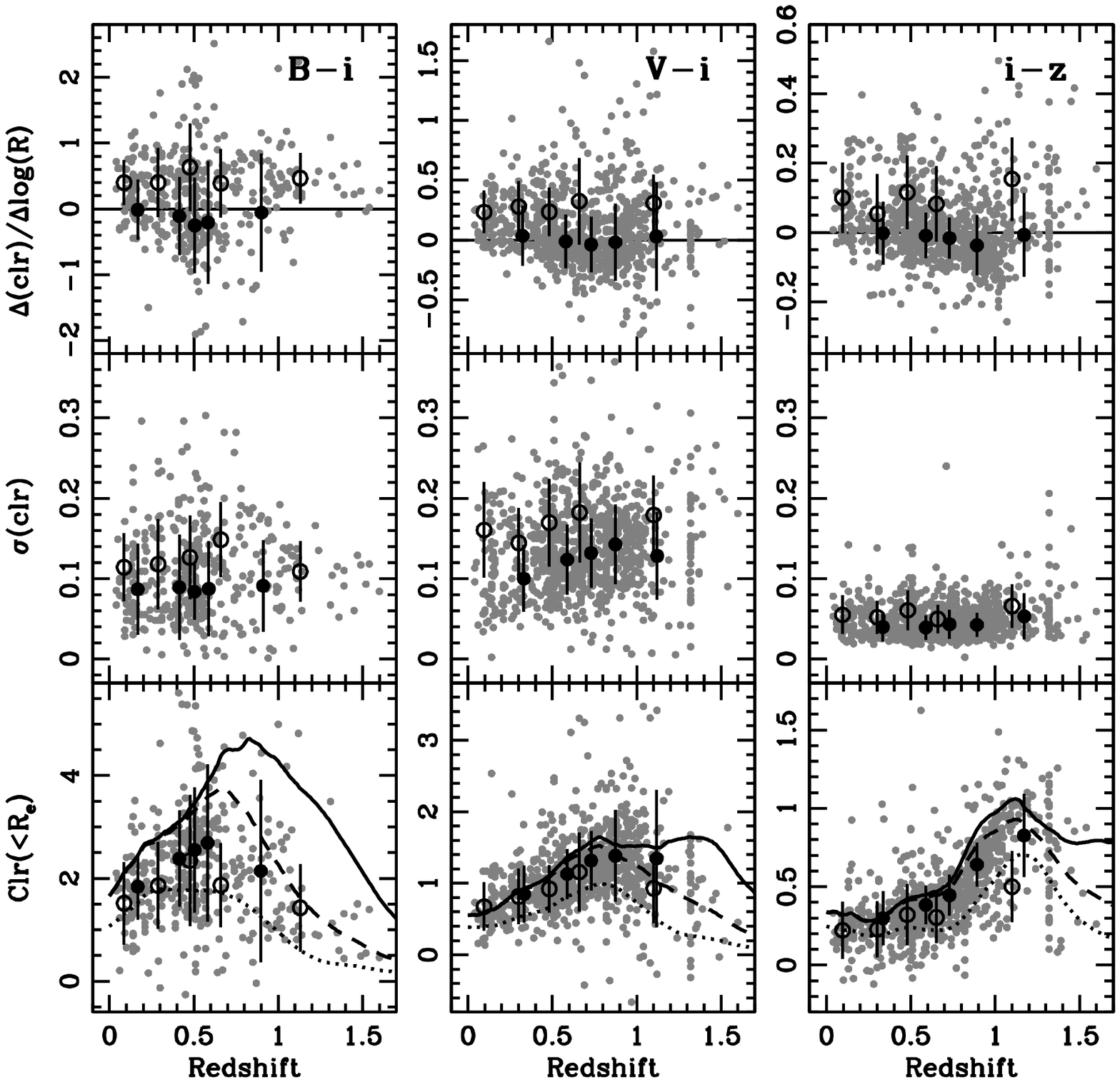}
\end{center}
\caption{{\sl Left:} The observer-frame colour gradient ({\sl top}),
scatter ({\sl middle}), and integrated colour within the half-light
radius ({\sl bottom}) are shown with respect to three observables:
redshift ({\sl left}); apparent half-light radius ({\sl middle}) and
$i_{\rm F775W}$ apparent magnitude ({\sl right}).  The full sample is
shown in grey. Black points correspond to average values of subsamples
(after a 5$\sigma$ clipping) binned at constant number of galaxies per
bin, and separated with respect to the best fit populations: old
(filled dots; photometric type t$\leq$ 2) and young (hollow dots;
t$>$2).  {\sl Right:} Redshift evolution of the intrinsic distribution
of the three available colours, as labelled. In both figures the lines
in the bottom panels correspond to the evolution of a set of
exponentially decaying star formation histories with solar
metallicity, started at z$_{\rm F}$=5 with a decay timescale of 0.5
(solid line) 1 (dashed line) and 8~Gyr (dotted line).
\label{fig:Clrobs}}
\end{figure*}

Regarding stellar mass, we choose a fixed reference mass
($10^{11}M_\odot$; thick solid line).  The sample is split in 1~dex
bins above and below this reference (dashed lines). For reference, the
dotted line tracks the characteristic stellar mass scale of
\citet{fon06}. A typical error bar is shown in both panels. The
vertical lines in both panels mark the redshift bins, with the full
range for a complete volume-limited sample given by the hatched
vertical regions.

Figure~\ref{fig:LFevol} shows the evolution in comoving number density
with respect to absolute magnitude ({\sl top}) or stellar mass ({\sl
bottom}) obtained from the previous diagram. The error bars are
estimated from a sum in quadrature of the Poisson noise in each bin
and the uncertainty in the absolute magnitude or mass. This
uncertainty is computed using a Monte Carlo algorithm in which noise
is added to the absolute magnitude or stellar mass of each
galaxy. 5,000 realizations are generated and each is analysed in the same
way as our original sample. The error bars give the
standard deviation within the ensemble. The evolution of the number
density can be modelled by a simple power law $n\propto (1+z)^\alpha$.
In luminosity there is no significant difference between the
``bright'' and ``faint'' subsamples. However, the most massive
galaxies show a remarkable weak trend with redshift, with a slope
significantly different to less massive galaxies. With the adopted fits to 
the data, an extrapolation to lower redshift yields a 
decrease of a factor $\sim 50$ in the number density of early-type
galaxies between z=0 and z=1 for the mass range $10^{10}-10^{11}M_\odot$;
whereas the decrease is just a factor 2 for galaxies with stellar
masses between $10^{11}$ and $10^{12}M_\odot$. This result is presented
in more detail in a letter \citep{fer08} where we show that this result
is in agreement with recent work in other surveys \cite[e.g.][]{con07} and with
recent semi-analytic models of galaxy formation \cite[e.g.][]{ks06}.

\section{Intrinsic Colour distribution} 

The {\sl internal} colour distribution of a galaxy can reveal
important information regarding its star formation and assembly
history. Early-type galaxies have smooth surface brightness
distributions, which make them ideal targets for this sort of study in
GOODS. For each colour map three pieces of information can be gathered
in a robust way: the average colour within an aperture (the projected
half-light radius in our case); the radial gradient of the colour
distribution, and the scatter about the best fit (assumed to be linear
between colour and $\log R$). We follow the same technique as
described in \citet{fl05}, briefly outlined here. 

\begin{figure}
\begin{center}
\includegraphics[width=3.5in]{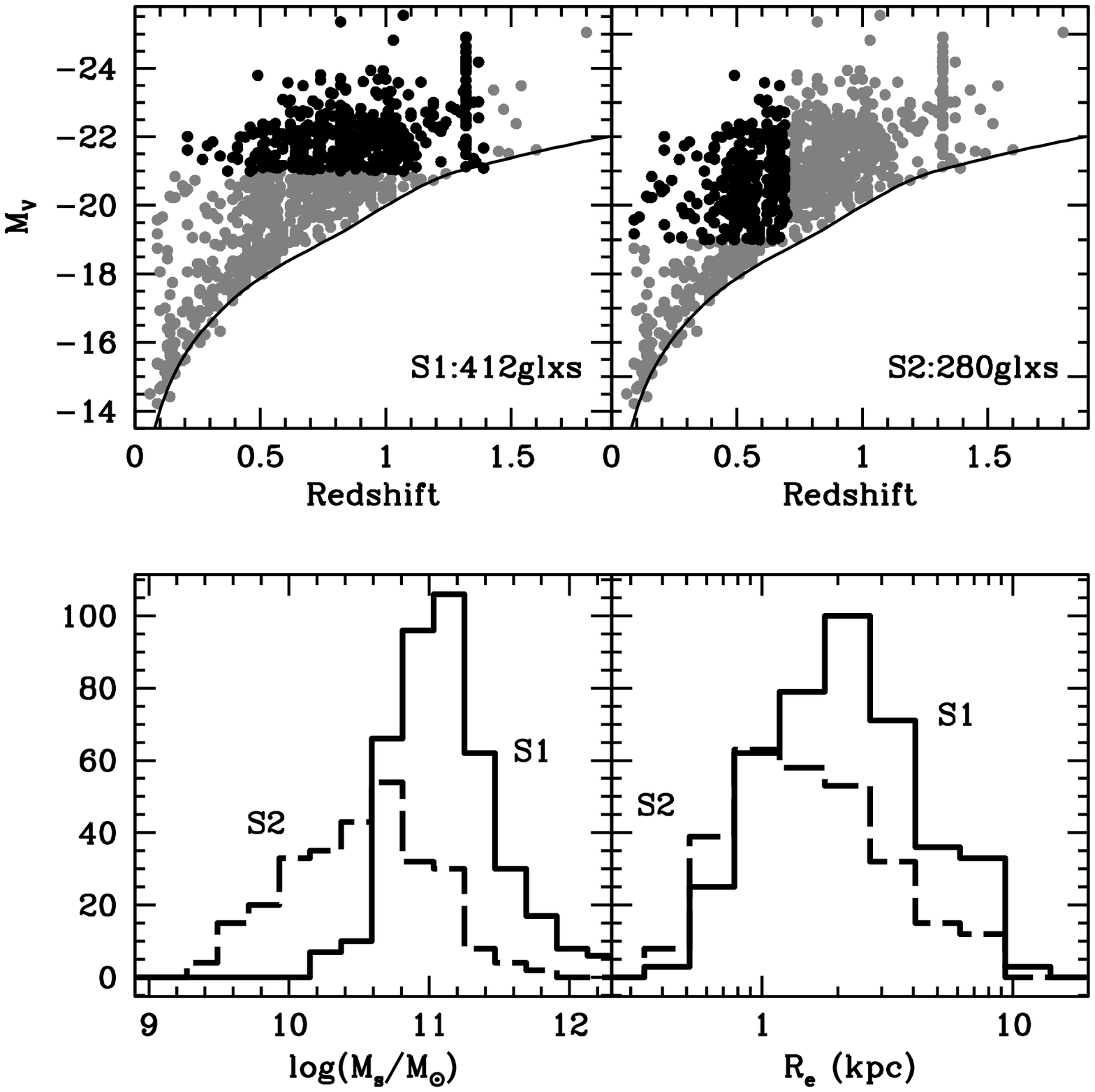}
\end{center}
\vskip-0.3truein
\caption{Volume limited subsamples of early-type galaxies
in GOODS N+S fields. S1 focuses on massive systems out to
redshift z$<$1.4, whereas S2 limits the redshift out to
z$<$0.7 in order to maximise the range of masses/luminosities.
The top panels show $V$-band absolute magnitude vs redshift, with
the $i_{\rm F775W}<24$ flux limit shown as a solid line.
The bottom panels show the histograms of each subsample in 
stellar mass and projected size.
\label{fig:VolLim}}
\end{figure}

The colour map is generated by convolving the image taken in one
filter with the Point Spread Function (PSF) of the other filter used
to define the colour. This is the most robust way to reduce the colour
trends caused by a different resolution between images in different
bandpasses. We used a PSF generated from stellar images in the same
dataset. \citet{fl05} includes an appendix that shows the --
negliglible -- difference in the colour gradients and scatter when
using either a stellar image from the same data or a synthetic PSF
generated by TinyTim\cite{tiny}.

The images span a wide range of signal-to-noise ratios on a
pixel-by-pixel basis. Hence, it is best to bin the pixels although
preserving spatial information as much as possible. A Voronoi
tessellation method was adopted, in which a target S/N is specified so
that the algorithm merges nearby pixels into tiles, keeping these
tiles as round as possible, following the algorithm by
\citet{ovt}. The process is finished when a target S/N {\sl per tile}
is reached. In \citet{fl05} we showed that estimates of radial colour
profiles are much improved as the outer parts of the galaxy have a
significantly lower noise after a Voronoi tessellation. We impose a
target S/N=10 per tile for each colour. The radial distribution of
colour vs $\log R/R_e$ is fit to a straight line. We also determine
the colour scatter about this best fit by using a biweight estimate
\citep{beers90}.

The left panel of figure~\ref{fig:Clrobs} shows the {\sl observer
frame} $V-i$ colour distribution as a function of redshift, half-light
radius and apparent $i$ band magnitude of the whole sample (grey
dots). The larger dots correspond to binning the sample at a fixed
number of galaxies per bin, separating them in photometric type
between red (t$\leq$ 2; solid circles) and blue (t$>$2; hollow
dots). Error bars represent the standard deviation within each
bin. For comparison, the lines at the bottom-left corner track
exponentially decaying star formation histories at solar metallicity
with formation redshift z$_{\rm F}$=5 and timescales $\tau=0.5$~Gyr
(solid), $1$~Gyr (dashed) and $8$~Gyr (dotted).

In the top panels, the radial colour gradient is shown. Galaxies with
red (blue) cores have negative (positive) gradients. We obtain a
consistent result with respect to \citet{fl05}. Galaxies with blue
cores are almost always photometrically classified as ``blue'' and
galaxies with red cores are classified as ``red''.  This is not a
trivial statement, as the overall colour of the galaxy need not
correlate with the radial gradient. The scatter does not correlate
with redshift, but there is a clear trend with size or apparent
magnitude.  This can be caused by a combination of physical and
passband-shifting effects and we defer the discussion to a later
figure.

On the right side of figure~\ref{fig:Clrobs}, a similar plot is
presented for the three available colours from the ACS images. The
$B-i$, $V-i$ and $i-z$ colour distributions are shown with respect to
redshift.  The lines in the bottom panels correspond to the same
models as in the left side of the figure. The trend is consistent
with all three colours, with blue galaxies having positive colour
gradients.

In order to extract meaningful trends of the intrinsic colour
distributions we have to define volume-limited subsamples that will
remove biases inherent to flux-limited
samples. Figure~\ref{fig:VolLim} illustrates our selection process.
The complete sample is shown in absolute magnitude vs redshift
space. The distribution shows the characteristic boundary caused by
the $i_{\rm F775W}=24$ limit. The solid line tracks a $i_{\rm
F775W}=24$ galaxy as a function of redshift, assuming a typical star
formation history for an early-type galaxy (photometric type t=1).  A
compromise has to be made when generating volume-limited samples:
either we probe as deep as possible in redshift -- cutting in absolute
magnitude, and thereby sacrificing faint galaxies, or we probe as
faint as possible the luminosity function -- cutting in redshift. In
order to explore these two choices we select two subsamples, shown as
black dots in each panel: S1 (412 galaxies; {\sl left}) is a sample
targeted to explore the redshift evolution, and is restricted to
galaxies brighter than $M_V<-21$. We also apply a cut in redshift
(z$<$1.4). Subsample S2 (280 galaxies; {\sl right}) targets a wide
range of luminosities out to z$<$0.7. In this case the volume-limited
sample extends to fainter galaxies $M_V<-19$.  The bottom panels of
figure~\ref{fig:VolLim} show the histograms in stellar mass ({\sl
left}) and projected physical size of each subsample.

\begin{figure}
\begin{center}
\includegraphics[width=3.5in]{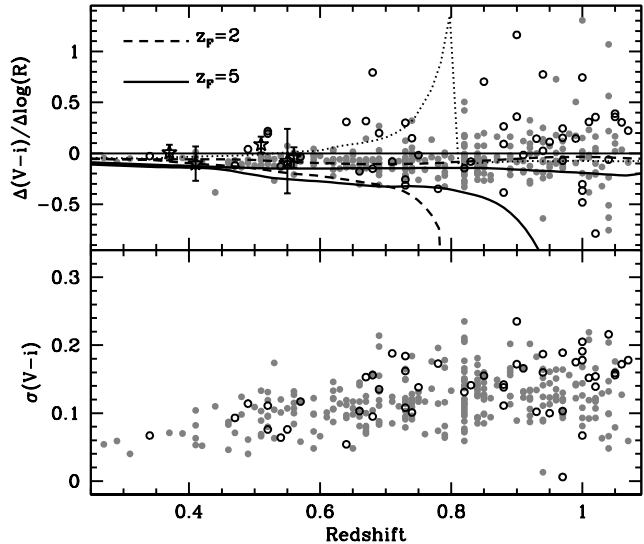}
\end{center}
\vskip-0.3truein
\caption{Intrinsic $V-i$ colour distribution of subsample S1: The slope
({\sl top}) and scatter of the radial colour gradient is shown 
with respect to redshift. The sample is separated into photometric
early- (grey dots) and late-types (open circles). Stars are gradient
measurements in a sample of cluster early-types from \citet{toh00}.
Two alternative models for the evolution of the colour gradient are shown,
depending on whether the gradient is caused by metallicity (thin, nearly
horizontal lines) or age (thick, curved lines). Two formation redshifts
are considered for each case: z$_{\rm F}$=5 (solid lines) or $2$ (dashed lines).
The dotted line corresponds to a metallicity sequence with formation
redshift z$_{\rm F}$=5, adding to the core 10\% (in mass) of a stellar
population formed at z=0.8.
\label{fig:clrS1}}
\end{figure}

Figure~\ref{fig:clrS1} focuses on the redshift evolution of the colour
distribution and shows the observer-frame $V-i$ colour gradient ({\sl
top}) and scatter of subsample S1. For comparison, the colour
gradients observed in a sample of moderate redshift clusters by
\citet{toh00} is shown as stars. Our sample is shown with the usual
coding with respect to photometric type: red (grey dots) and blue
(open circles). In the appendix of \cite{fl05} simulations of galaxies
with the same characteristics as the GOODS galaxies were done to show that
from photometric errors one could expect a typical colour scatter 
$\sigma(V-i)\simlt 0.03$~mag, with a maximum of 0.05~mag for the faintest
galaxies. Figure 15 in the appendix of \cite{fl05} shows that both
the observed scatter and slope are intrinsic.

We include in the figure a simple model prediction which uses the
local $B-R$ colour gradients of elliptical galaxies in the local
Universe as constraint \citep{pel90}.  The set of two nearly
horizontal lines -- just below the $\Delta (V-i)/\Delta\log(R)=0$ axis
-- assumes the z=0 intrinsic colour gradient to be caused by a pure
metallicity sequence, i.e. we fix an age throughout the galaxy and
compute the difference in metallicity that will give the observed
colour gradient in local early-type galaxies. Then we evolve this
model backwards in time to assess the evolution in the colour
gradient.  Standard models with exponentially decaying star formation
histories are used for the fit. The timescale is fixed at $\tau=1$~Gyr
and the formation epoch is either z$_{\rm F}$=2 (solid line) or $5$
(dashed line).  The second sets of models -- which curve down at
z$<$0.6 -- assume a pure age sequence for the colour gradient at fixed
(solar) metallicity, with a similar set of models (in this case the
bluer outer region is modelled by a slightly longer star formation
timescale). As expected, this model gets very steep with increasing
redshift because of the strong dependence of colour with age. A third
model (dotted line) overlays on top of a metallicity sequence formed
at z${\rm F}$=5 a second, younger stellar population {\sl in the
core}, formed at z=0.8. This younger component contributes only 10\%
in mass, illustrating the effect that a small episode caused by a
minor merger would have on the colour gradient.

This result illustrates that blue cores have a radial range of ages,
being youngest at their centres. The large blue colour gradients
observed in many of the blue cores cannot be explained by an
``inverted'' gradient in metallicity, i.e. more metal-rich populations
outwards. If we were to explain the large range in colour between the
inner and the outer parts of the galaxy by metallicity alone, one
would have to invoke unrealisticaly low metallicities at the galaxy
centre. Furthermore, this scenario would predict no redshift
dependence on the fractional contribution of blue cores, in contrast
to the observations. In this subsample (S1) of luminous ($M_V<-21$)
early-type galaxies, the median redshift of blue and red early-types
is 1.10 and 0.85, respectively, i.e. bluer early-type galaxies (at
least at the massive end) contribute more significantly at z$\simgt$1.

\begin{figure}
\begin{center}
\includegraphics[width=3.5in]{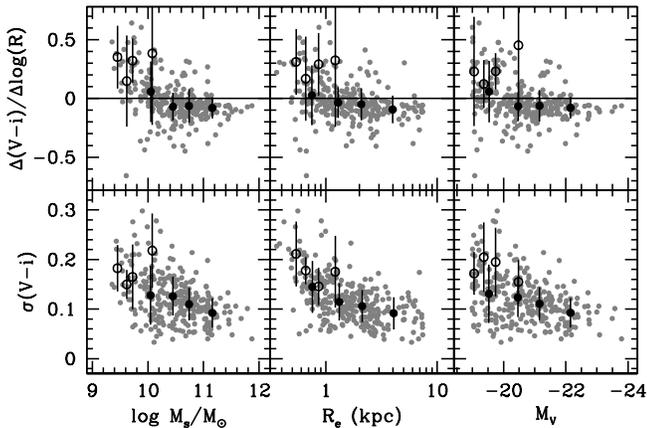}
\end{center}
\vskip-0.3truein
\caption{Intrinsic ($V-i$) colour distribution of subsample S2: The slope
({\sl top}) and scatter of the observed $V-i$ colour distribution --
roughly rest-frame $B-V$ over the redshift range of S2 -- is shown
with respect to stellar mass ({\sl left}); projected half-light radius
({\sl middle}) and $V$-band absolute magnitude ({\sl right}). 
The black points with error bars correspond to
average values of subsamples separated with respect to photometric
redshift type: old (filled dots; type t$\leq$ 2) and young (hollow
dots; t$>$2). 1$\sigma$ error bars are shown (after a 5$\sigma$
clipping).  The binning is done at constant number of galaxies per
bin.
\label{fig:clrS2}}
\end{figure}

The relation between colour distribution and galaxy properties is
explored in figure~\ref{fig:clrS2}; where the colour gradient ({\sl
top}) and scatter is shown with respect to stellar mass; projected
half-light radius and $V$-band absolute magnitude for subsample
S2. Blue cores typically have stellar masses below $10^{10}M_\odot$
and half-light radii $R_e\simlt 1$~kpc, a signature of downsizing. In
this subsample the average stellar mass of the blue ellipticals is
$5\cdot 10^9 M_\odot$ compared to $4\cdot 10^{10}M_\odot$ for red
ellipticals. This strong mass-dependence suggests that AGN activity
alone is not enough to explain the blue cores \citep{men05}.

\section{The Kormendy Relation}

The Kormendy relation \citep[hereafter KR;][]{kr77} is a tight linear
correlation between average surface brightness -- typically measured
within the half-light radius ($R_e$) -- and the logarithm of $R_e$.
It is one of the projections of the Fundamental Plane \citep{dd87},
which relates surface brightness, size and velocity dispersion.  The
Fundamental Plane and its projections are a consequence of the
dynamics of early-type galaxies and their underlying stellar
populations.  KR is therefore the ``next best thing'' to the
Fundamental Plane when measurements of velocity dispersion are not
available. The observed dispersion of the KR at z=0 is $\sim
0.4$~mag\,arcsec$^{-2}$\citep{lab03}. The redshift evolution of the
Kormendy relation can be used to constrain the assembly and
formation history. A significant change in slope with redshift
can reveal the presence of star formation or the rearrangement of
the stars because of ``dynamical activity'' (i.e. mergers).

\begin{figure}
\begin{center}
\includegraphics[width=3.5in]{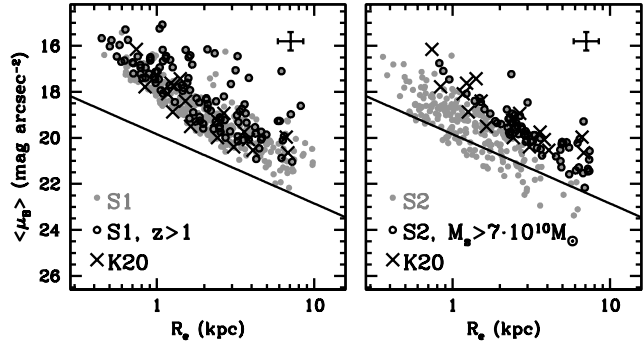}
\end{center}
\vskip-0.3truein
\caption{Kormendy relation of samples S1 ({\sl left}) and S2 ({\sl
right}) in the rest-frame $B$-band (grey dots). Open circles dots are the
upper quartile of the redshift (S1) or the stellar mass
distribution (S2), as labelled. A typical error bar is shown. The data
have been corrected for the $(1+z)^4$ cosmological dimming. The line
represents the local relation observed in the Coma cluster \citep{jfk95}. 
The 17 galaxies from the K20 sample are shown as crosses \citep{k20}.
\label{fig:Kor}}
\end{figure}

Figure~\ref{fig:Kor} shows the Kormendy relation of both
volume-limited samples S1 and S2 (grey dots). The surface brightness
is computed as the average within the half-light radius.  Open black
dots correspond to the upper quartile in redshift (S1, {\sl left}) or
stellar mass (S2, {\sl right}), as labelled. The solid line is the
local relation observed in the Coma cluster \citep{jfk95} and the
crosses are a sample of 17 galaxies from the K20 survey \citep{k20}.
Sample S1 -- which extends to higher redshift than S2 -- has a
significantly brighter zero point, caused by the effect of lookback
time on the stellar populations. For comparison, a simple stellar
population formed at z$_{\rm F}$=5 and solar metallicity undergoes a
fading of $\sim 1$~mag in the rest-frame $B$ band between redshifts
z=1 and $0$, similar to the offset found in the left panel of
figure~\ref{fig:Kor}. Sample S2 ({\sl right}) has a KR compatible with
local early-type galaxies (again with a significant shift of the zero
point caused by lookback time). As expected, galaxies with a higher
stellar mass content appear brighter than less massive galaxies with
the same size, roughly a direct consequence of a non-homologous
surface brightness distribution \citep{gg03}. An estimate of the
typical error bar is shown in each panel. For the surface brightness
we use the scatter of the K-corrections applied to obtain $B$-band
magnitudes (0.4~mag). The true error should be most certainly smaller
than this K-correction.  The error in the size is taken from the worst
case scenario when retriving half-light radii from simulations (see
appendix). The K20 sample from \citet{k20} (crosses) span a narrow
redshift range ($0.9\cdots 1.3$), and are compatible with our
high-redshift ({\sl left}) or massive ({\sl right}) subsamples.

Notice the major difference between the selection criterion of this
work and \citet{fl05}. In that paper, a Kormendy selection was {\sl
imposed} on the visually classified sample, so that early-types that
were found not to evolve into the local KR were rejected from the
sample. In that case, a large number of low-redshift, blue early-types
were rejected. It was discussed whether misclassification of bright
knots of star formation or an active nucleus could explain this
rejection.  In this paper we use visual classification as the only
selection criterion. We later define volume-limited subsamples. In
this process we end up rejecting similar type of galaxies as in
\citet{fl05} {\sl for different reasons}. Figure~\ref{fig:volz} shows
that most of the faint, low-redshift galaxies in our sample are
removed from S1 or S2 to enforce completeness. Those are mainly blue
early-types with small sizes. The addition of those galaxies
to S1 or S2 would result in significant scatter towards the
bottom-left corner of the Kormendy relation in figure~\ref{fig:Kor},
in a similar fashion to the presence of the ``dwarf galaxy branch''
presented in the Kormendy Relation plot of \citet{capa92}. Hence we
conclude that a significant fraction of the galaxies rejected in
\citet{fl05}  because of the KR constraint must be dwarf ellipticals at
z$\simlt$ 0.4.

\begin{figure}
\begin{center}
\includegraphics[width=3.5in]{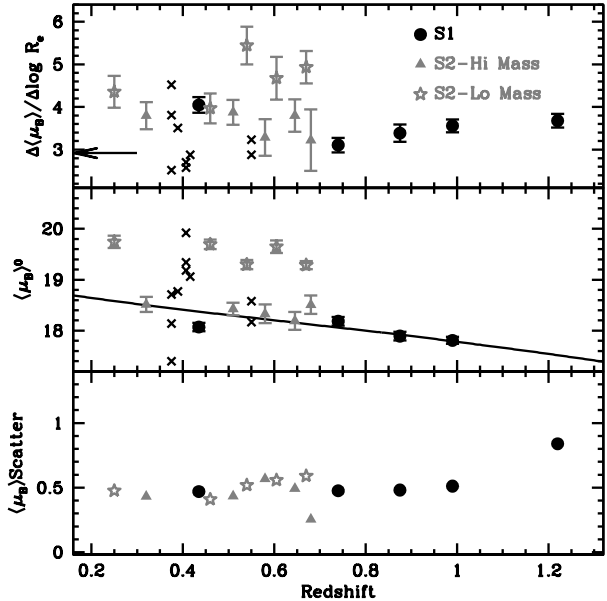}
\end{center}
\vskip-0.3truein
\caption{Redshift evolution of the Kormendy relation. Sample S1 is shown
as black circles, sample S2 (grey) has been split with respect to the median
stellar mass within that sample ($3.5\cdot 10^{10}M_\odot$). The bins
in redshift are chosen such that the number of galaxies per bin is
constant for each set.  A Kormendy relation is fit, and this plot
gives the slope ({\sl top}), zero point ({\sl middle}; in mag\,arcsec$^{-2}$) 
and biweight scatter about the best fit ({\sl bottom}; in mag\,arcsec$^{-2}$). 
The arrow in the top panel is the z$\sim$0 slope obtained
by \citet{lab03} and the crosses are the cluster observations of
\citet{zieg99}. The line in the middle panel corresponds to the
evolution of an exponentially decaying star formation rate started at
z$_{\rm F}$=5, with timescale $\tau = 0.5$~Gyr and solar metallicity.
\label{fig:Korz}}
\end{figure}

Figure~\ref{fig:Korz} shows the redshift evolution of the KR of
subsamples S1 and S2 in terms of slope ({\sl top}); average surface
brightness ({\sl middle}) and scatter ({\sl bottom}; using a biweight
estimator). Sample S2 is split at the median value of stellar mass
($3.5\cdot 10^{10}M_\odot$). The fit is done with a standard least
squares algorithm, binning the samples at constant number of galaxies
per bin (hence the unequal separation between dots in the figure). The
arrow in the top panel represents the z=0 slope of the Kormendy
Relation \citep{lab03}. Sample S1 (black dots) -- focussing on bright
galaxies out to the highest redshifts -- shows a remarkable agreement
with the local observations, and with moderate redshift clusters
\citep{zieg99}. We extend the conclusion of \citet{zieg99} of a
negliglible change in the slope of the KR to a wide range of redshifts
and to a field environment. The most massive galaxies in sample S2
(grey triangles) also agree with the local observations, whereas a
selection of the galaxies with the lower masses in S2 (grey stars)
have larger scatter and steeper slopes.

In the middle panel we compare the observed trend of the average
surface brightness with the fading of a typical old population as
labelled in the figure caption.  Notice that the least massive
galaxies of sample S2 are consistently brighter than this estimate,
reflecting the presence of younger stellar populations. It is
interesting to notice that the $\sim 0.5$~mag arcsec$^{-2}$ scatter
about the best fit ({\sl bottom}) does not change significantly with
redshift or sample.

To summarise, we find the Kormendy relation of $z\sim 0.3-1.2$ to be
the same as at $z\sim 0$. The slope and scatter of the correlation
does not change and the zero point, i.e. the typical surface
brightness changes in agreement with pure passive evolution. Only
lower mass spheroidals (M$_s<3.5\cdot 10^{10}$M$_\odot$) show a
significant departure from the average $z=0$ Kormendy relation,
appearing brighter in $\langle\mu_B\rangle^0$ and with a steeper
slope. We believe this result is caused by a combination of younger
stellar populations and a significantly different surface brightness
profile compared to more massive early-type galaxies, similar to the
$z=0$ case \citep{gg03}.

\section{Summary and Conclusions}

We present a sample of visually-classified early-type galaxies over
the 360~arcmin$^2$ area covered by the ACS/{\sl HST} images of the Great
Observatories Origins Deep Survey \citep[GOODS,][]{goods}, including both the North
and South fields. The classification is only limited in flux ($i_{\rm
AB}<24$), i.e. no colour cuts or surface brightness profile
constraints are imposed in the selection. The sample comprises 910
early-type galaxies, for which we compile spectroscopic quality redshifts
from publicly available datasets for 541 galaxies (i.e. $\sim 60$\% of the
total sample). For the rest we estimate photometric redshifts from
available data (both {\sl HST} and ground-based). We find the comoving
number density to depend strongly on stellar mass, such that the number
of galaxies with $M_s\simgt 10^{11}M_\odot$ only increases a factor 2
between z=1 and z=0, whereas the number of galaxies with stellar
mass $10^{10}<M_s/M_\odot <10^{11}$ increase a factor $\sim 50$ in the
same redshift range. In the full, flux-limited sample, we find 26\%
of our early-types are blue (i.e. corresponding to a star formation
history with timescale longer that 1~Gyr, or with a significant 
amount of young stars). This fraction is similar to the
number of low-redshift early-types with a significant amount of recent
star formation \citep{Kav07}.

The depth and superb resolution of the ACS images enable us to explore
the {\sl intrinsic} colour distribution. We perform a technique based
on a Voronoi tessellation \citep[already applied in ][]{fl05} that
allows a robust determination of the radial colour gradient and
scatter. In figure~\ref{fig:Clrobs} we show that the colour gradients
are strongly dependent on the overall photometric type, such that red
galaxies have mostly red cores and blue early-type galaxies have blue
cores. This is a non-trivial statement, as one could imagine star
formation in the outer parts of the galaxy, resulting in blue galaxies
with red cores.  We find central star formation in otherwise red
galaxies is very rare.  On the other hand, the majority of galaxies
with (recent) star formation exhibit most of this star formation in
their centers. These observations suggest an inside-out formation
process in these galaxies, ruling out recent attempts at explaining
elliptical galaxy formation with an outside-in formation scenario
\citep{pip06}. An inside-out scenario implies that regardless of the
way star formation is triggered -- primordial collapse versus
merging/interactions, gas is accumulated towards the centre, starting
star formation in the central regions, subsequently ``diffusing''
outwards. One could argue whether the mechanism explaining these blue
cores reduces to minor mergers \citep{Ellis01,Kav07b}. Robust
confirmation of the actual cause of these blue cores will require the
analysis of abundance ratio gradients \citep[see e.g.][]{FS02}.

Figure~\ref{fig:clrS1} shows that the local observations of radial
colour gradients in elliptical galaxies \cite[e.g. ][]{pel90} mostly
correspond to a range of metallicities. This can be throught of as the
{\sl intrinsic} equivalent of the pure metallicity sequence of
\citet{ka97}, used in that paper to explain the colour-magnitude
relation {\sl among} galaxies. Our results are in agreement with the
sample of cluster galaxies at low and moderate redshift 
\citep{toh00,lab04}, but our
sample increases by an order of magnitude the number and accuracy of
the colour gradient estimates. \citet{lab04} only find optical colour
gradients compatible with zero slope. We do find non-zero, albeit small,
optical gradients and a significant trend with stellar mass, such that
the most massive galaxies have almost always red cores.

Defining volume-limited samples, we find that blue galaxies only
contribute 4\% to the total number of bright ($M_V<-21$) early-type
galaxies out to z$<$1.4. Most of the blue early-type galaxies in our
-- visually classified -- sample correspond to low-mass systems, and
are removed when taking volume-limited samples. At face value, we
could state that significant star formation in early-type galaxies
only takes place for galaxies less massive than $M_s\sim
10^{10}M_\odot$ even out to redshifts z$\sim$1.  As shown by
\citet{bell07}, star formation in a substantial fraction of blue cloud
galaxies needs to be shut off between z=1 and z=0, making them evolve
to the red sequence, thereby explaining the significant growth in mass
there. However, in addition to considering the contribution with
respect to mass and star formation history, it is also well known that
blue cloud and red sequence differ morphologically, with spiral and
irregular galaxies dominating the blue cloud whereas red sequence
galaxies are mostly early-types. Since our sample of early-types was
selected on pure morphological grounds, those early-types with
(recent) star formation confirm the scenario of \citet{bell07}, in
that these objects will not only evolve onto the red sequence in terms
of colour, but also in terms of morphology.

The Kormendy relation was explored in two volume-limited samples and a
remarkable consistency is found with the KR at z=0, with no
significant change of the slope of the Kormendy relation out to 
z$\sim$1.2. 
The scatter about the best fit does not change significantly either.
Only the average surface brightness is found to change in the
Kormendy relation, with an evolution consistent with pure passive
evolution of old stellar populations. 

\section*{Acknowledgments}
We thank Bahram Mobasher for useful comments about the paper, and 
Alister Graham for providing us with his code to compute
correction factors for magnitudes and radii.  IF acknowledges the
Nuffield foundation for partial support of this project. T.L.\ is
supported within the framework of the Excellence Initiative by the
German Research Foundation (DFG) through the Heidelberg Graduate
School of Fundamental Physics (grant number GSC 129/1). This paper is
based on observations made with the NASA ESA \emph{Hubble Space
Telescope}, obtained from the data archive at the Space Telescope
Science Institute, which is operated by the Association of
Universities for Research in Astronomy, Inc. under NASA contract NAS
5-26555. The NIR observations have been carried out using the Very
Large Telescope at the ESO Paranal Observatory under Program
ID:LP168.A-0485


\clearpage
\onecolumn

\appendix
\section{Extraction of sizes and magnitudes from the ACS images}

As discussed in \S\S 2.1 it is important to correct for the loss of
light expected for a typical surface brightness distribution. In order
to estimate the accuracy of the retrieved size, apparent magnitude and
surface brightness we ran a number of simulations, generating postage
stamps with the same noise characteristics as the ones taken from the
ACS images. The simulations correspond to smooth distributions of
galaxies with a Sersic profile \citep{ser68}. The simulations
realistically map the error in each pixel using the weight maps of the
ACS images to determine a correlation between surface brightness per
pixel and S/N.

For a given choice of apparent magnitude and half-light radius we
generate 501$\times$501 stamps with the same pixel size as the ACS
images (0.03~arcsec). For each galaxy in the ensemble we take a random
value for the Sersic index (between $n=2$ and $4$); ellipticity
(between $e=0$ and $0.6$) and position angle. A total of 16 ensembles are
generated for four different values of apparent magnitude 
($i_{\rm F775W}=\{18, 20, 22, 24\}$) and half-light radius 
($R_e=\{5, 8, 10, 20\}$ pixels). 
Each model galaxy is convolved by the PSF of ACS in the $i$ band (obtained
from a set of stellar images as discussed in \S4) and is passed
through the same code as the original sample to determine the
half-light radius, magnitude and surface brightness.

\begin{figure}
\begin{center}
\includegraphics[width=4in]{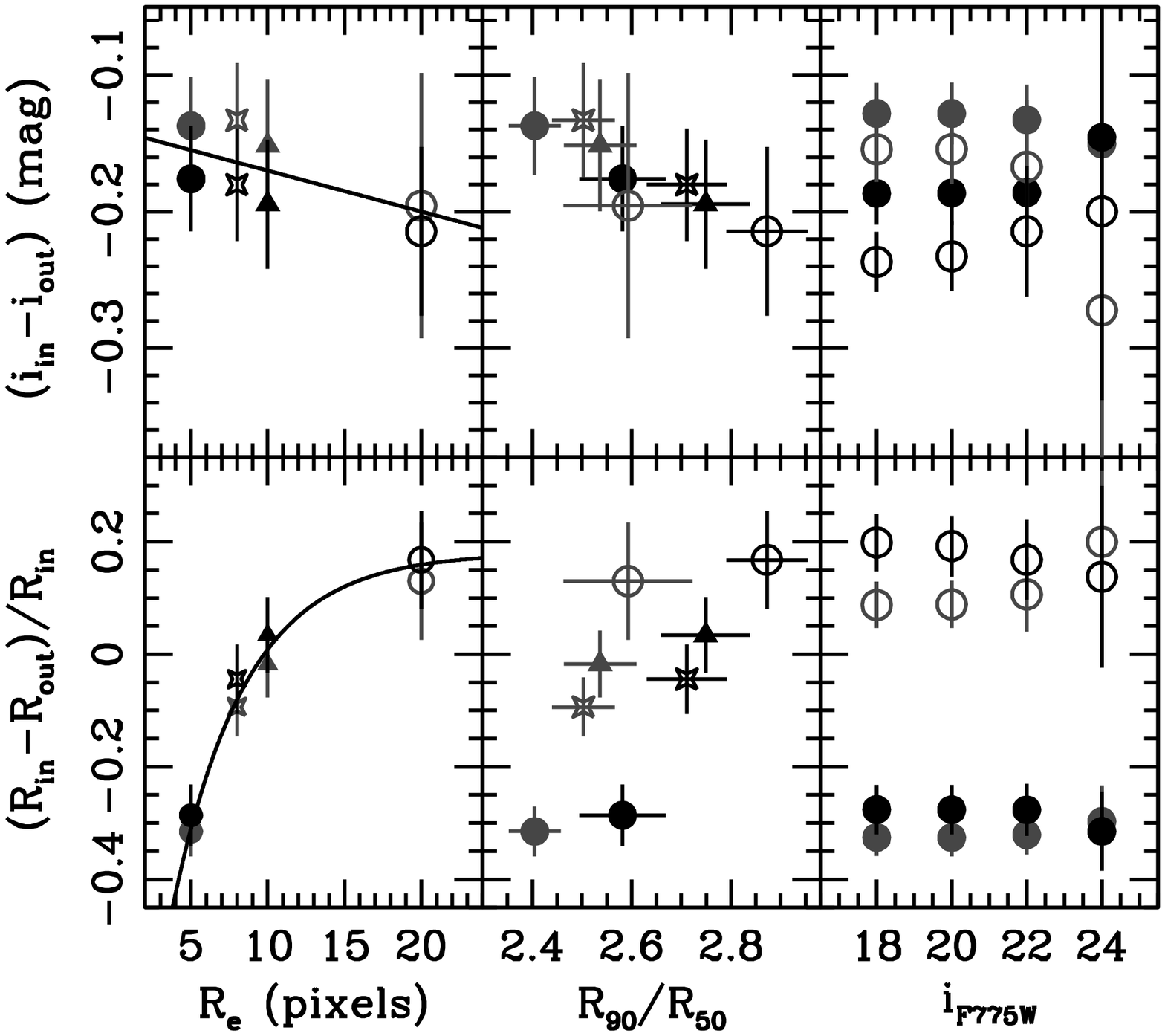}
\end{center}
\caption{Recovered $i_{\rm F775W}$ (top), and half-light radius
(bottom) of simulated galaxies, $1\sigma$ error bars shown. Each
symbol corresponds to a choice of $R_e$. The sample of 8,000 simulated
galaxies is separated with respect to R$_{90}$/R$_{50}$, such that the
grey (black) dots represent the subsample below (above) the median. The
highest discrepancy in the recovery of the half-light radius occurs
for values close to the FWHM of the PSF ($\sim 3$ pixels). The
simulations are shown with respect to $Re$ ({\sl left}), concentration
{\sl middle} -- measured as R$_{90}$/R$_{50}$ -- and apparent
magnitude ({\sl right}).  The lines in the leftmost panels give the
corrections used for flux and half-light radius (equations
\ref{eq:corr1} and \ref{eq:corr2}) at the median value of concentration.
Notice that the panels on the right only show the subsamples with
the largest and smallest value of $R_e$ to avoid crowding the figure.
\label{fig:ReSim1}}
\end{figure}

Figure~\ref{fig:ReSim1} shows the result (1$\sigma$ error bars). Notice
the systematic trend in the offset with respect to half-light radius
({\sl left}) or concentration (measured as R$_{90}$/R$_{50}$; {\sl
right}). The figure also shows that the correction should depend to
first order on $R_e$, whereas concentration is a second-order correction.
The total sample of 8,000 galaxies is split about the median value of
R$_{90}$/R$_{50}$, showing the most concentrated galaxies in black.
Our results are consistent with \citet{gra05b}, as
concentration correlates with Sersic index. However our regime
of observations is different from their work, as our galaxies
have sizes significantly closer to the PSF.

\begin{figure}
\begin{center}
\includegraphics[width=4in]{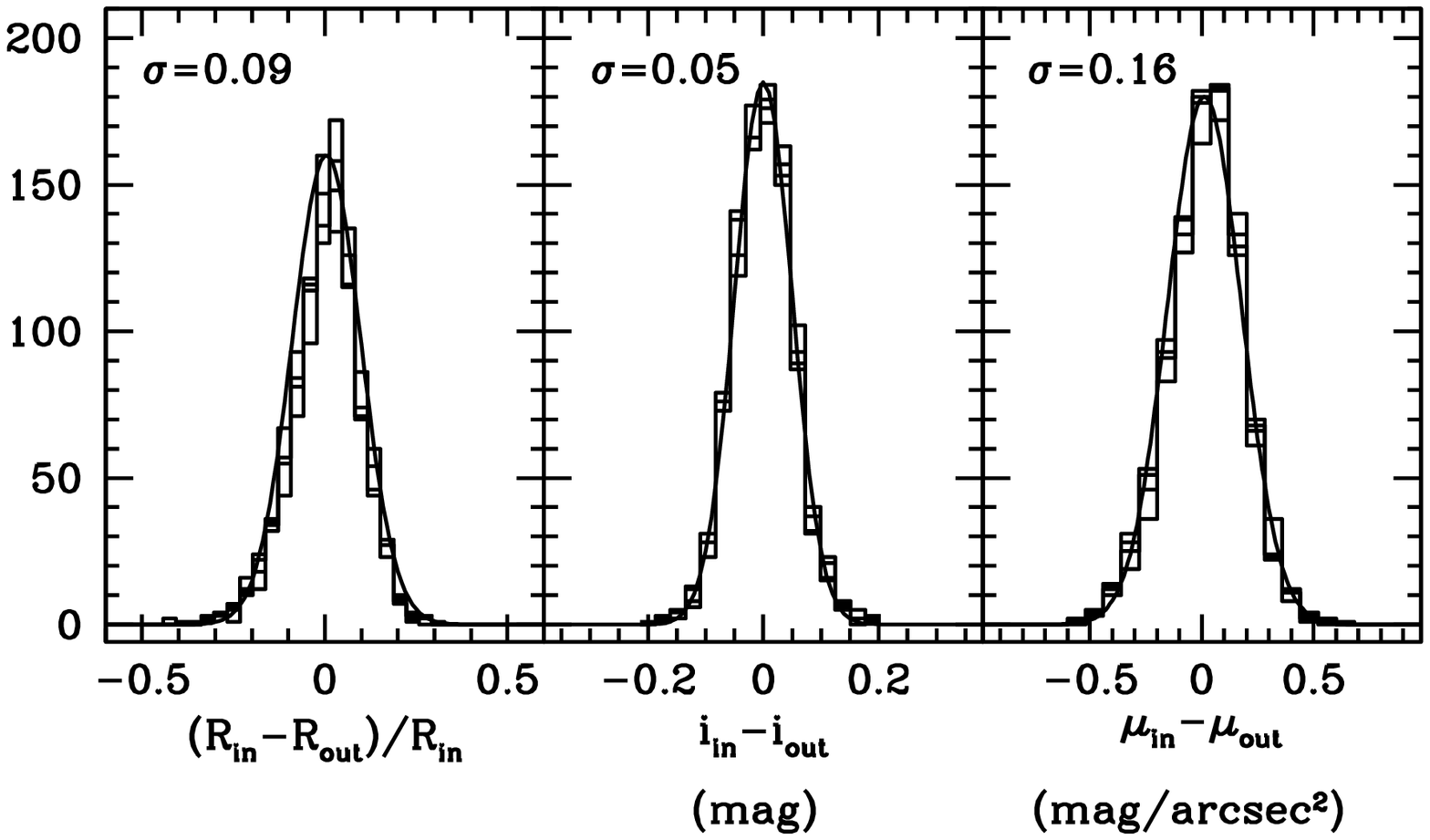}
\end{center}
\caption{Recovery of four simulated samples of 910 galaxies with
the same distribution in apparent magnitude and size as the original
sample. The histograms compare the original and the recovered size ({\sl left});
magnitude ({\sl middle}) and apparent surface brightness 
along with a Gaussian that characterizes the uncertainty for the complete
sample.
\label{fig:ReSim2}}
\end{figure}

The simulations shown in figure~\ref{fig:ReSim1} motivate a simple
correction dominated by R$_e$ with concentration modelled as a 
linear change to the zero point of the correction. The corrections
-- shown in the leftmost panels of figure~\ref{fig:ReSim1} for
the median value of R$_{90}$/R$_{50}$ -- are:

\be 
R_{\rm out} =
R_{\rm in}\Big( 0.57+1.5e^{-0.2R_{\rm in}}-0.244\frac{R_{90}}{R_{50}}\Big),
\label{eq:corr1}
\ee 

\noindent
and 

\be 
i_{\rm out} = i_{\rm in} + 0.05 + 0.003R_{\rm in}+0.068\frac{R_{90}}{R_{50}},
\label{eq:corr2}
\ee 

\noindent
with the radii given in pixels. These two expressions are used to make
a correction on apparent magnitude and half-light radius.  In order to
assess the accuracy after the corrections we generate four samples of
910 simulated galaxies with the same apparent magnitude and size as
our observed sample. The Sersic index, ellipticity and position angle
are chosen randomly in the same way as described above.
Figure~\ref{fig:ReSim2} shows the histograms comparing recovered and
original size, magnitude and apparent surface brightness after
applying the corrections given by equations~\ref{eq:corr1} and
\ref{eq:corr2}.  The lines are Gaussian fits to the data, with the
standard deviation shown in the top left corner of each panel. These
are our {\sl official} uncertainties, namely 9\% in $R_e$, 0.05~mag
in the {\sl TOTAL} $i_{\rm F775W}$ magnitude and 0.16~mag/arcsec$^2$ in
$i_{\rm F775W}$ apparent surface brightness.

\section{Recovering stellar masses}

\begin{figure}
\begin{center}
\includegraphics[width=4in]{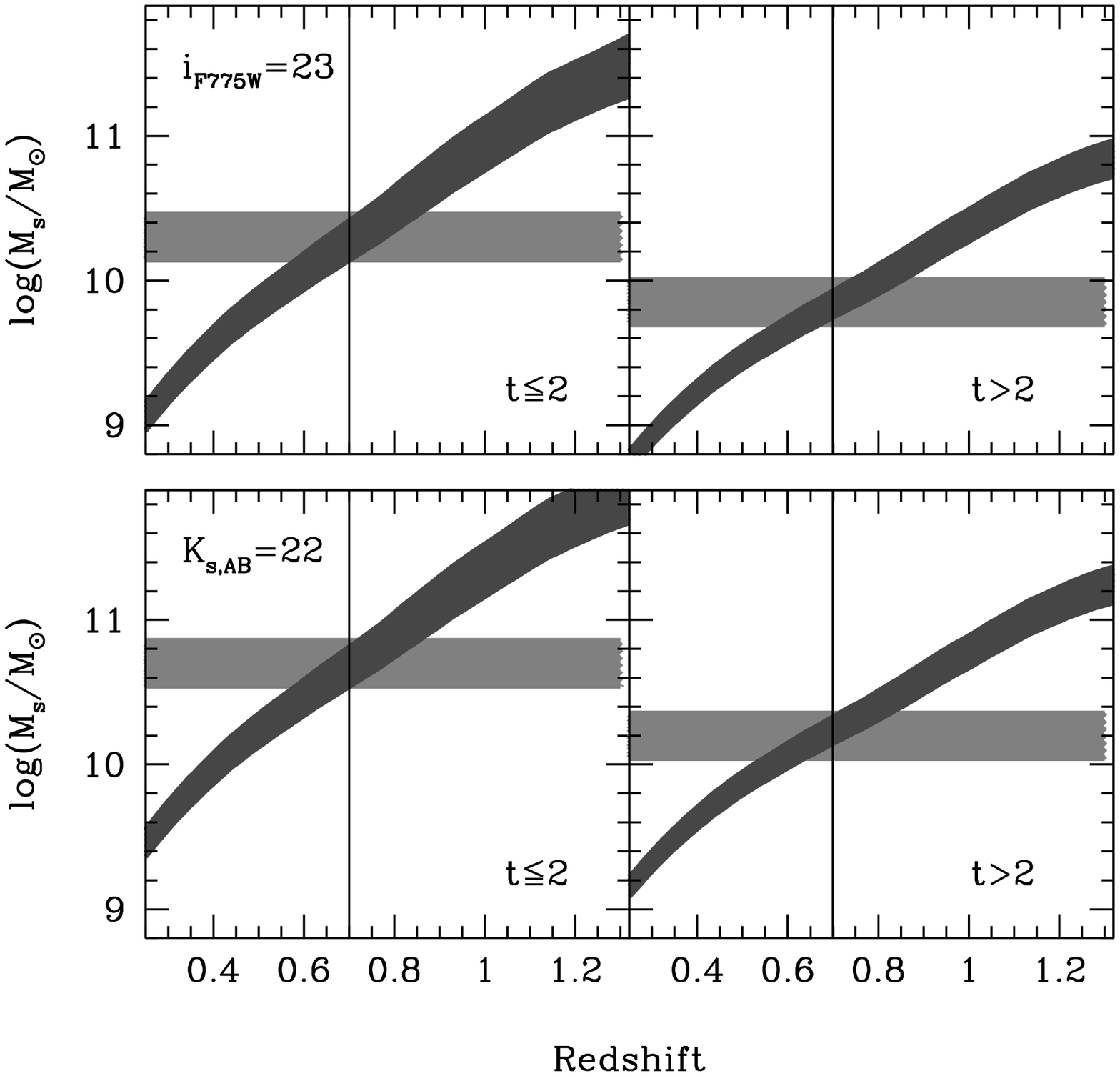}
\end{center}
\caption{Comparing the accuracy of stellar mass estimates
using $i_{\rm F775W}$ ({\sl top}) vs a more ``traditional''
method with NIR magnitudes ({\sl bottom}). A fiducial apparent
magnitude is assumed as labelled. The dark grey region gives the
stellar mass expected as a function of redshift for photometrically
``red'' ({\sl left}) and ``blue'' galaxies. To guide the eye, we
assume a z=0.7 galaxy (vertical line). The light gray horizontal
shading represents a 0.3~dex uncertainty about the predicted value.
Notice both $i_{\rm F775W}$ and $K_s$ give similar accuracy over the
redshift probed by our sample.
\label{fig:logM1}}
\end{figure}

The stellar masses presented in this paper are extracted from the
total apparent $i_{\rm F775W}$ magnitudes. One could argue that NIR
photometry might be a more reliable tracer of the stellar mass. To
confirm the validity of $i_{\rm F775W}$ photometry {\sl over the redshift
range probed by our sample}, we show in figure~\ref{fig:logM1} the
predicted mass for a $i_{\rm F775W}=23$ (AB) galaxy ({\sl top}) or a $K_s=22$
(AB) galaxy ({\sl bottom}). The result is shown according to
photometric type, with the left (right) panels showing red (blue)
galaxies. The models are identical to those described in \S\S2.3.  The
dark grey shading gives the expected stellar mass as a function of redshift,
and -- to guide the eye -- we include a vertical line at z=0.7 (roughly the
median redshift of the sample).  As
expected, for a later (i.e. bluer) photometric type, the stellar mass
is smaller. One should emphasize here that within our volume-limited
subsamples S1 and S2, the fraction of galaxies corresponding to the
panels on the right amount only to 4 and 11\%, respectively.

The horizonal light grey shading shows a 0.3~dex uncertainty in stellar
mass for a z=0.7 galaxy. Notice that in this redshift range $i_{\rm F775W}$
is as accurate a tracer of stellar mass as $K_s$, notwithstanding the fact
that the high resolution of the ACS images allows us to determine more
accurately the TOTAL apparent magnitude of the galaxies.

As a second check, we retrieved the publicly available ISAAC/{\sl VLT}
$K_s$ images of the GOODS/CDFS field (Retzlaff et al., in preparation). The images
were analysed with SExtractor \citep{sex} using the provided zero point
and weight maps. The catalogue of detected sources was cross-correlated
with our sample, and only those galaxies with a SExtractor {\tt flag}=0
were selected for the comparison. We used MAG\_BEST and further applied
the same correction to obtain total magnitudes as in appendix~A. 
Figure~\ref{fig:logM2} shows the comparison, along with the equivalent 
Gaussian distribution, which has a mean of 
$\langle\log M_s(K_s)-\log M_s(i)\rangle=-0.04$~dex and standard deviation
$0.13$~dex. The mean and standard deviation are reduced to $-0.02$ and $0.11$~dex,
respectively when the sample is restricted to $i_{\rm F775W}<23$ galaxies.
We conclude that our $i$-band extracted stellar masses are accurate to
within the quoted $0.3$~dex uncertainty (including population synthesis
effects).

\begin{figure}
\begin{center}
\includegraphics[width=4in]{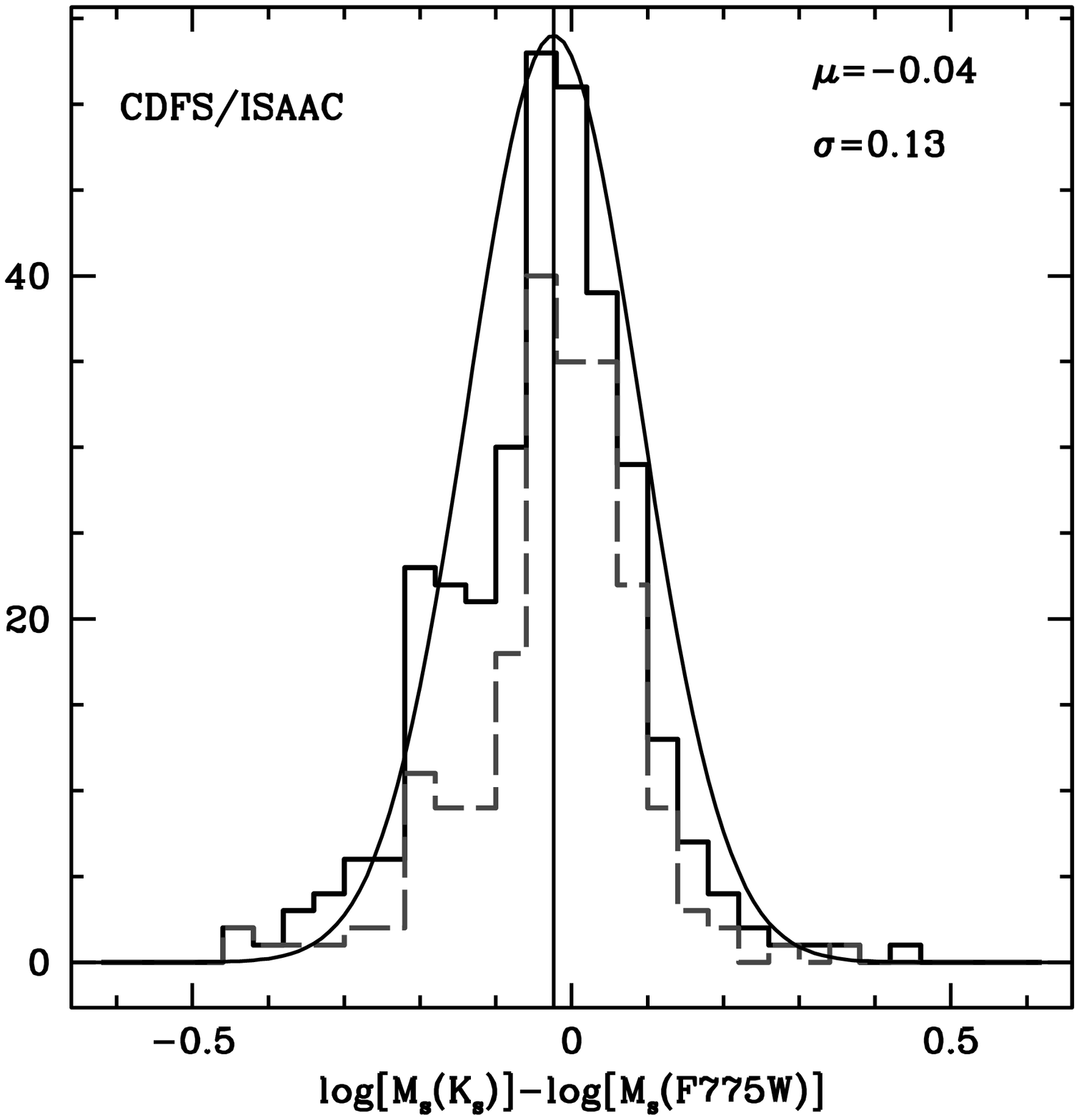}
\end{center}
\caption{Comparison between stellar masses from this work -- extracted from the
{\sl corrected} F775W magnitudes -- and NIR masses using the $K_s$ magnitudes
of the CDFS using the ISAAC images (Retzlaff et~al., in preparation). 
Identical flux corrections
are applied to the NIR magnitudes as those obtained for F775W (see appendix A).
The grey histogram corresponds to a subsample of the brighter 
($i_{\rm F775W}<23$) galaxies, which give an offset of $\mu=-0.02$ and a standard
deviation of $\sigma=0.11$~dex.
\label{fig:logM2}}
\end{figure}

\end{document}